\documentclass[preprint,3p,times,onecolumn]{elsarticle}
\pdfoutput=1
\usepackage{etoolbox}
\usepackage[utf8]{inputenc}
\usepackage[colorlinks]{hyperref}
\usepackage{graphicx}
\usepackage{natbib}
\usepackage{tabularx}
\usepackage{lipsum}
\usepackage{wrapfig}
\usepackage{amsmath}
\usepackage{float}
\usepackage{subfigure}
\usepackage{gensymb}
\usepackage{hyperref}
\usepackage{cleveref}
\usepackage{chngcntr}
\usepackage{multirow}
\usepackage{makecell}
\usepackage{array}
\usepackage{booktabs}
\usepackage[none]{hyphenat} 
\usepackage{enumitem}
\usepackage{ragged2e}
\usepackage{amssymb}
\usepackage[section]{placeins}
\usepackage{siunitx}
\usepackage{epigraph}
\usepackage{lineno}
\usepackage{braket}
\usepackage{xparse,nameref}
\usepackage[usestackEOL]{stackengine}
\usepackage{xcolor}
\usepackage{mathtools}
\usepackage{multirow}
\usepackage{makecell}
\usepackage{array}
\usepackage{booktabs}
\usepackage{textcase}
\usepackage[toc,page]{appendix}
\usepackage{pdfpages}
\usepackage{titlesec}

\titleformat{\subsection} {\normalfont\bfseries}{\thesubsection}{1em}{}
\titleformat{\paragraph} {\normalfont\bfseries\itshape}{\paragraph}{1em}{}
\usepackage{dcolumn}   
\newcommand\Bstrut{\rule[-1.3ex]{0pt}{0pt}} 

\newcommand\TstrutLarge{\rule{0pt}{3.6ex}}       
\newcommand\BstrutLarge{\rule[-2.3ex]{0pt}{0pt}} 

\usepackage[tablename=TABLE, labelsep=period, figurename=Figure, skip=0pt, font=small]{caption}

\makeatletter
\def\ps@pprintTitle{%
 \let\@oddhead\@empty
 \let\@evenhead\@empty
 \def\@oddfoot{}%
 \let\@evenfoot\@oddfoot}
\makeatother

\begin{document}
  \begin{frontmatter}
    \title{\textbf{Correlated and Integrated Directionality for sub-MeV solar neutrinos in Borexino}}
\author[London1,Munchen]{M.~Agostini}
\author[Munchen]{K.~Altenm\"{u}ller}
\author[Munchen]{S.~Appel}
\author[Kurchatov]{V.~Atroshchenko}
\author[Juelich]{Z.~Bagdasarian\fnref{Berkeley}}
\author[Milano]{D.~Basilico}
\author[Milano]{G.~Bellini}
\author[PrincetonChemEng]{J.~Benziger}
\author[LNGS]{R.~Biondi}
\author[Milano]{D.~Bravo\fnref{Madrid}}
\author[Milano]{B.~Caccianiga}
\author[Princeton]{F.~Calaprice}
\author[Genova]{A.~Caminata}
\author[Virginia]{P.~Cavalcante\fnref{LNGSG}}
\author[Lomonosov]{A.~Chepurnov}
\author[Milano]{D.~D'Angelo}
\author[Genova]{S.~Davini}
\author[Peters]{A.~Derbin}
\author[LNGS]{A.~Di Giacinto}
\author[LNGS]{V.~Di Marcello}
\author[Princeton]{X.F.~Ding}
\author[Princeton]{A.~Di Ludovico} 
\author[Genova]{L.~Di Noto}
\author[Peters]{I.~Drachnev}
\author[Dubna,Milano]{A.~Formozov}
\author[APC]{D.~Franco}
\author[Princeton,GSSI]{C.~Galbiati}
\author[LNGS]{C.~Ghiano}
\author[Milano]{M.~Giammarchi}
\author[Princeton]{A.~Goretti\fnref{LNGSG}}
\author[Juelich,RWTH]{A.S.~G\"ottel}
\author[Lomonosov,Dubna]{M.~Gromov}
\author[Mainz]{D.~Guffanti}
\author[LNGS]{Aldo~Ianni}
\author[Princeton]{Andrea~Ianni}
\author[Krakow]{A.~Jany}
\author[Munchen]{D.~Jeschke}
\author[Kiev]{V.~Kobychev}
\author[London,Atomki]{G.~Korga}
\author[Juelich,RWTH]{S.~Kumaran}
\author[LNGS]{M.~Laubenstein}
\author[Kurchatov,Kurchatovb]{E.~Litvinovich}
\author[Milano]{P.~Lombardi}
\author[Peters]{I.~Lomskaya}
\author[Juelich,RWTH]{L.~Ludhova}
\author[Kurchatov]{G.~Lukyanchenko}
\author[Kurchatov]{L.~Lukyanchenko}
\author[Kurchatov,Kurchatovb]{I.~Machulin}
\author[Mainz]{J.~Martyn}
\author[Milano]{E.~Meroni}
\author[Dresda]{M.~Meyer}
\author[Milano]{L.~Miramonti}
\author[Krakow]{M.~Misiaszek}
\author[Peters]{V.~Muratova}
\author[Munchen]{B.~Neumair}
\author[Mainz]{M.~Nieslony}
\author[Kurchatov,Kurchatovb]{R.~Nugmanov}
\author[Munchen]{L.~Oberauer}
\author[Mainz]{V.~Orekhov}
\author[Perugia]{F.~Ortica}
\author[Genova]{M.~Pallavicini}
\author[Munchen]{L.~Papp}
\author[Juelich,RWTH]{L.~Pelicci}
\author[Juelich]{\"O.~Penek}
\author[Princeton]{L.~Pietrofaccia}
\author[Peters]{N.~Pilipenko}
\author[UMass]{A.~Pocar}
\author[Kurchatov]{G.~Raikov}
\author[LNGS]{M.T.~Ranalli}
\author[Milano]{G.~Ranucci}
\author[LNGS]{A.~Razeto}
\author[Milano]{A.~Re}
\author[Juelich,RWTH]{M.~Redchuk\fnref{Padova}}
\author[Perugia]{A.~Romani}
\author[LNGS]{N.~Rossi}
\author[Munchen]{S.~Sch\"onert}
\author[Peters]{D.~Semenov}
\author[Juelich]{G.~Settanta\fnref{ISPRA}}
\author[Kurchatov,Kurchatovb]{M.~Skorokhvatov}
\author[Juelich,RWTH]{A.~Singhal}
\author[Dubna]{O.~Smirnov}
\author[Dubna]{A.~Sotnikov}
\author[LNGS,Kurchatov]{Y.~Suvorov\fnref{Napoli}}
\author[LNGS]{R.~Tartaglia}
\author[Genova]{G.~Testera}
\author[Dresda]{J.~Thurn}
\author[Peters]{E.~Unzhakov}
\author[Dubna]{A.~Vishneva}
\author[Virginia]{R.B.~Vogelaar}
\author[Munchen]{F.~von~Feilitzsch}
\author[GSI,Juelich,RWTH]{A.~Wessel}
\author[Krakow]{M.~Wojcik}
\author[Hamburg]{B.~Wonsak}
\author[Mainz]{M.~Wurm}
\author[Genova]{S.~Zavatarelli}
\author[Dresda]{K.~Zuber}
\author[Krakow]{G.~Zuzel}

\fntext[Berkeley]{Present address: University of California, Berkeley, Department of Physics, CA 94720, Berkeley, USA}
\fntext[Napoli]{Present address: Dipartimento di Fisica, Universit\`a degli Studi Federico II e INFN, 80126 Napoli, Italy}
\fntext[Madrid]{Present address: Universidad Autónoma de Madrid, Ciudad Universitaria de Cantoblanco, 28049 Madrid, Spain}
\fntext[LNGSG]{Present address: INFN Laboratori Nazionali del Gran Sasso, 67010 Assergi (AQ), Italy}
\fntext[Padova]{Present address: Dipartimento di Fisica e Astronomia dell’Università di Padova and INFN Sezione di
Padova, Padova, Italy}
\fntext[ISPRA]{Present address: Istituto Superiore per la Protezione e la Ricerca Ambientale, 00144 Roma, Italy}
\address{\bf{The Borexino Collaboration}}
\address[APC]{APC, Universit\'e de Paris, CNRS, Astroparticule et Cosmologie, Paris F-75013, France}
\address[Dubna]{Joint Institute for Nuclear Research, 141980 Dubna, Russia}
\address[Genova]{Dipartimento di Fisica, Universit\`a degli Studi e INFN, 16146 Genova, Italy}
\address[Krakow]{M.~Smoluchowski Institute of Physics, Jagiellonian University, 30348 Krakow, Poland}
\address[Kiev]{Institute for Nuclear
Research of NAS Ukraine, 03028 Kyiv, Ukraine}
\address[Kurchatov]{National Research Centre Kurchatov Institute, 123182 Moscow, Russia}
\address[Kurchatovb]{ National Research Nuclear University MEPhI (Moscow Engineering Physics Institute), 115409 Moscow, Russia}
\address[LNGS]{INFN Laboratori Nazionali del Gran Sasso, 67010 Assergi (AQ), Italy}
\address[Milano]{Dipartimento di Fisica, Universit\`a degli Studi e INFN, 20133 Milano, Italy}
\address[Perugia]{Dipartimento di Chimica, Biologia e Biotecnologie, Universit\`a degli Studi e INFN, 06123 Perugia, Italy}
\address[Peters]{St. Petersburg Nuclear Physics Institute NRC Kurchatov Institute, 188350 Gatchina, Russia}
\address[Princeton]{Physics Department, Princeton University, Princeton, NJ 08544, USA}
\address[PrincetonChemEng]{Chemical Engineering Department, Princeton University, Princeton, NJ 08544, USA}
\address[UMass]{Amherst Center for Fundamental Interactions and Physics Department, UMass, Amherst, MA 01003, USA}
\address[Virginia]{Physics Department, Virginia Polytechnic Institute and State University, Blacksburg, VA 24061, USA}
\address[Munchen]{Physik-Department, Technische Universit\"at  M\"unchen, 85748 Garching, Germany}
\address[Lomonosov]{Lomonosov Moscow State University Skobeltsyn Institute of Nuclear Physics, 119234 Moscow, Russia}
\address[GSSI]{Gran Sasso Science Institute, 67100 L'Aquila, Italy}
\address[Dresda]{Department of Physics, Technische Universit\"at Dresden, 01062 Dresden, Germany}
\address[Mainz]{Institute of Physics and Cluster of Excellence PRISMA+, Johannes Gutenberg-Universit\"at Mainz, 55099 Mainz, Germany}
\address[GSI]{GSI Helmholtzzentrum f\"ur Schwerionenforschung, Planckstrasse 1, D-64291 Darmstadt, Germany}
\address[Juelich]{Institut f\"ur Kernphysik, Forschungszentrum J\"ulich, 52425 J\"ulich, Germany}
\address[RWTH]{III. Physikalisches Institut B, RWTH Aachen University, 52062 Aachen, Germany}
\address[London]{Department of Physics, School of Engineering, Physical and Mathematical Sciences, Royal Holloway, University of London, Egham, TW20 OEX, UK}
\address[London1]{Department of Physics and Astronomy, University College London, London, UK}
\address[Atomki]{Institute of Nuclear Research (Atomki), Debrecen, Hungary}
\address[Hamburg]{University of Hamburg, Institute of Experimental Physics, Luruper Chaussee 149, 22761 Hamburg, Germany}

\begin{abstract}
Liquid scintillator detectors play a central role in the detection of neutrinos from various sources. In particular, it is the only technique used so far for the precision spectroscopy of sub-MeV solar neutrinos, as demonstrated by the Borexino experiment at the Gran Sasso National Laboratory in Italy. The benefit of a high light yield, and thus a low energy threshold and a good energy resolution, comes at the cost of the directional information featured by water Cherenkov detectors, measuring $^8$B solar neutrinos above a few MeV. In this paper we provide the first directionality measurement of sub-MeV solar neutrinos which exploits the correlation between the first few detected photons in each event and the known position of the Sun for each event. This is also the first signature of directionality in neutrinos elastically scattering off electrons in a liquid scintillator target.  This measurement exploits the sub-dominant, fast Cherenkov light emission that precedes the dominant yet slower scintillation light signal. Through this measurement, we have also been able to extract the rate of $^{7}$Be solar neutrinos in Borexino. The demonstration of directional sensitivity in a traditional liquid scintillator target paves the way for the possible exploitation of the Cherenkov light signal in future kton-scale experiments using liquid scintillator targets. Directionality is important for background suppression as well as the disentanglement of signals from various sources.
     \end{abstract}
       
    \end{frontmatter}
    \twocolumn 

\section{Introduction}

Present-day solar neutrino detectors either use scintillation light~\cite{detector-paper, KamLAND} or Cherenkov light~\cite{super-kamiokande_IV, SNO} for neutrino detection. In a liquid scintillator (LS) detector like Borexino, when a neutrino scatters off an electron, this recoil electron excites the LS molecules, which in turn emit isotropic scintillation light with a wavelength distribution and time profile that depends on the LS. In water Cherenkov neutrino detectors such as Super-Kamiokande~\cite{super-kamiokande_IV} and SNO~\cite{SNO}, the recoil electron scattered off by the neutrino produces Cherenkov light which can then be used for direction reconstruction, a powerful tool for background rejection and separation of different signals on an event-by-event basis. While Cherenkov detectors have an upper hand on the directional reconstruction of neutrinos, LS detectors benefit from a high light yield, a low detection threshold, and a good energy resolution. For example, in Borexino, scintillation provides a high effective light yield of about 500 photoelectrons at 1\,MeV with 2000 live PMTs~\cite{BxCalibPaper}, while in Super Kamiokande, the light yield is about 34 photoelectrons at 3.5\,MeV~\cite{super-kamiokande_IV}. It has to be noted that the low energy threshold of scintillator detectors is possible only with extremely high levels of radio-purity, which has been achieved in Borexino through the choice of detector materials~\cite{detector-paper} and special purification campaigns~\cite{phase2-nusol}.

In recent years, there has been an increased interest in developing techniques for the hybrid detection of scintillation and Cherenkov light with the goal to obtain a directional signature in scintillator detectors, as proposed by the THEIA experiment~\cite{THEIA}. Cherenkov light has a different wavelength distribution and its time profile peaks at earlier times relative to scintillation light. In traditional LS detectors, it is challenging to disentangle the sub-dominant Cherenkov light due to the timing overlap with the dominant scintillation light. This can be overcome with a hybrid detector which can help obtain particle direction through Cherenkov light, while preserving the excellent energy resolution and low energy threshold of a scintillator detector. A favorable Cherenkov-to-scintillation ratio can be achieved by tuning the scintillation time profile and wavelength distribution, for example with a slow LS~\cite{Wang-slowLS} or a water-based LS~\cite{wbls} or through the use of novel scintillation materials like quantum-dots~\cite{QDLS}. The CHESS experiment~\cite{CHESS_2} showed the separation of Cherenkov light with fast photo-sensors and reconstruction of Cherenkov rings in a LS. The potential use of fast photo-sensors has also been investigated for direction reconstruction in LS detectors in the context of double-beta decay~\cite{double-beta}. There is ongoing investigation of hardware for Cherenkov-scintillation separation through spectral sorting~\cite{bandpass}. The directionality of electron anti-neutrinos in LS has previously been used in the CHOOZ and Double CHOOZ experiments~\cite{CHOOZ_directionality, double_CHOOZ_directionality}. The inverse beta decay of electron anti-neutrinos produces a prompt positron signal and a delayed neutron signal. As the neutron is more likely to be emitted in the forward direction, the difference in the reconstructed positions between the positron and the neutron can be used to statistically determine the average anti-neutrino direction, given a sufficiently large number of events ~\cite{IBD_directionality}. While this method is useful for reactor electron anti-neutrinos, it is not applicable to solar neutrinos that interact via elastic scattering off electrons in the LS.

In this paper, we employ the novel \emph{Correlated and Integrated Directionality} (CID) method to measure Cherenkov signals of sub-MeV solar neutrinos in a conventional high light yield LS detector, without any specialized hardware or LS mixtures. Though developed for Borexino, other large-volume scintillator detectors like KamLAND~\cite{KamLAND}, JUNO~\cite{JUNO}, and SNO+~\cite{SNO+} can benefit from this analysis technique. This method is different from the event-by-event directional reconstruction using Cherenkov light mentioned previously and instead relies on the well-known position of the neutrino source, which in the case of solar neutrinos is the Sun. In this CID technique, we correlate the detected PMT hit pattern to the well-known position of the Sun, and then this is integrated over a large number of events. This procedure results in a distribution of the angle between the hit PMT and the solar neutrino direction, with respect to the reconstructed event vertex. Electrons scattered by solar neutrinos will produce a characteristic signature in the angular distribution in comparison to the isotropic radioactive background which is uncorrelated to the Sun. This signature makes it possible to disentangle and measure the solar neutrino signal through CID.

Section~\ref{sec:bxdet} of this paper describes the structure of the Borexino detector, the composition of the liquid scintillator and other detector characteristics. Section~\ref{sec:solar-nu} gives an overview of solar neutrinos and the relevant backgrounds for the directional analysis.  Section~\ref{sec:cumul_che} then explains in detail the \emph{Correlated and Integrated Directionality} (CID) approach used in this paper. The selection of the solar neutrino dataset, using a favorable energy region and fiducial volume for the CID analysis is described in Section~\ref{sec:data-sel}, while Section~\ref{sec:mc_prod} explains the properties of Cherenkov and scintillation light from the Borexino Monte Carlo (MC) simulation. The analysis strategy used for the directionality measurement is discussed in Section~\ref{sec:analysis}. To  minimize the systematic uncertainty of the effective Cherenkov group velocity, we have performed a Cherenkov calibration of the Borexino MC in Section~\ref{sec:gamma_calib} using the available gamma sources from the 2009 calibration campaign~\cite{BxCalibPaper}. The other systematic effects for the CID analysis are explored in Section~\ref{sec:sys-others}.  In Section~\ref{sec:results}, we then present the measurement of sub-MeV solar neutrinos using the CID method and conclude the paper in Section~\ref{sec:conclusion}.

\section{The Borexino experiment}
\label{sec:bxdet}

Borexino is an ultra radio-pure liquid scintillator detector~\cite{detector-paper}, located in Hall C of the Gran Sasso National Laboratory in central Italy at a depth of some 3800\,metres water equivalent, where the muon flux is suppressed by a factor of $\sim$10$^{6}$ with respect to sea level. The experiment started data-taking in May 2007 and ended in October 2021 and has three main Phases, namely Phase-I (May 2007\textendash May 2010)~\cite{phase1-nusol}, Phase-II (December 2011\textendash May 2016)~\cite{nature-phase2, phase2-nusol}, and Phase-III (July 2016\textendash October 2021)~\cite{CNO-nature}. The success of the Borexino experiment lies fundamentally in its unprecedented radio-purity which was achieved through the choice and
development of innovative methods and low background materials as well as due to a decade-long effort consisting of several purification cycles. Due to this, the $^{238}$U and $^{232}$Th contamination reached the levels of $< 9.4\times10^{-20}$g/g (95\% C.L.) and $<5.7\times10^{-19}$g/g (95\% C.L.), respectively~\cite{phase2-nusol}.

\begin{figure}[t]
    \centering
    \includegraphics[width=0.4\textwidth]{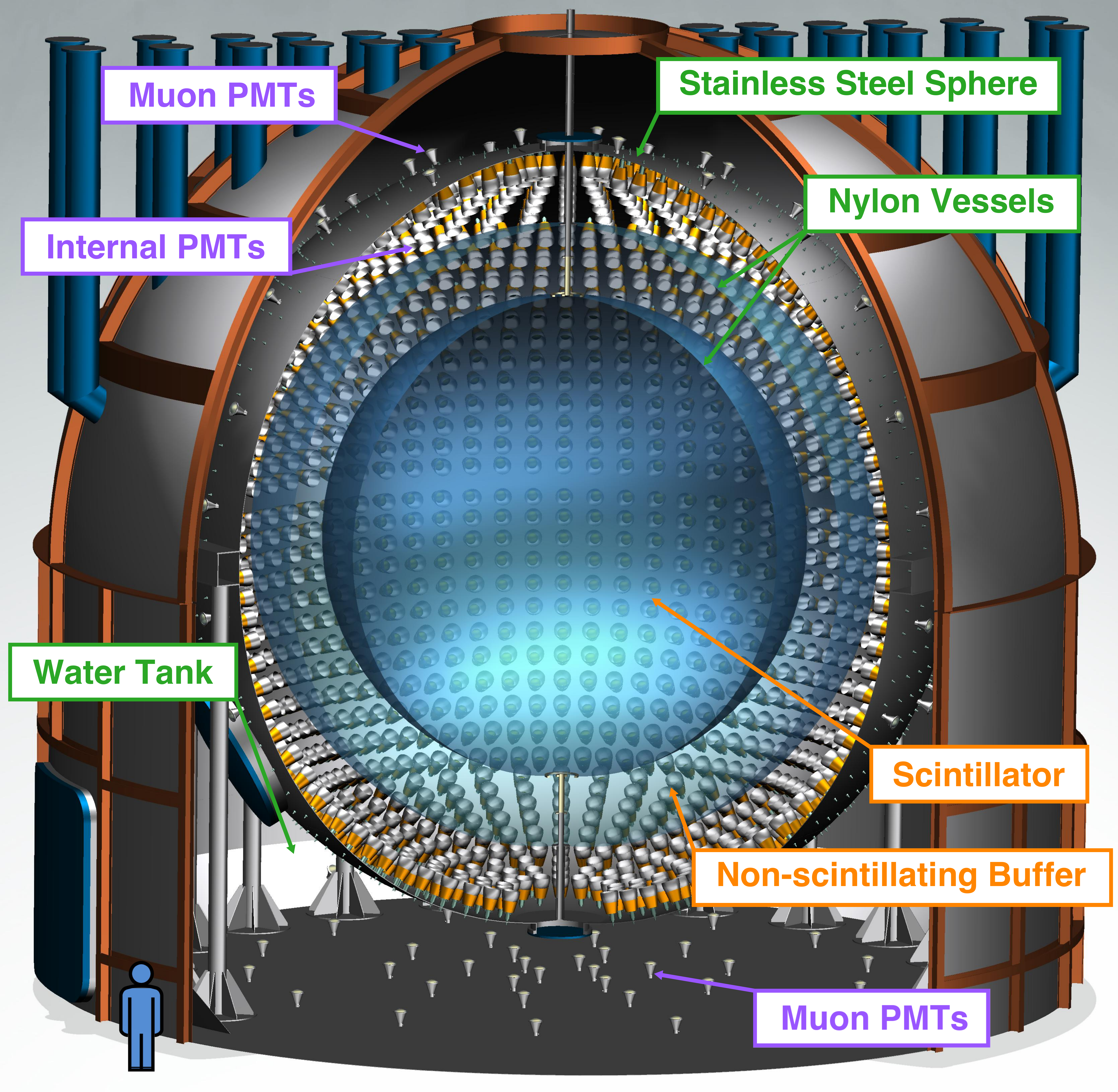}
    \caption{Schematic representation of the Borexino detector.}
    \label{fig:borex}
\end{figure}

The general scheme of the Borexino detector is shown in Figure~\ref{fig:borex}. The detector has a concentric multi-layer structure. The active medium in Borexino is an organic \emph{ liquid scintillator} (LS), with a nominal total mass of 280\,t, confined within a nylon \emph{Inner Vessel} (IV) of 4.25\,m radius. The scintillator is composed of pseudocumene (PC) solvent doped with a fluorescent dye PPO. The scintillator density is $(0.878 \pm  0.004)$\,g\,cm$^{-3}$~\cite{Geo-longpaper}. The shape of the IV changes with time, because of a small leak of the LS from the IV to the buffer region which started around April 2008~\cite{phase1-nusol, Geo-longpaper}. Therefore, its shape is reconstructed every week using the spatial distribution of the radioactive contaminants on its surface. The LS is surrounded by a non-scintillating buffer liquid (inner buffer). This buffer region is held by a nylon \emph{Outer Vessel} (OV) with a radius of 5.50\,m, followed by a second outer buffer region, which in turn is surrounded by a stainless steel sphere (SSS) with a radius of 6.85\,m.
Both regions between the IV/OV and OV/SSS are filled with PC doped with dimethyl phthalate (DMP) which acts as a quenching agent on PC.
The SSS holds 2212 8-inch photomultiplier tubes (PMTs), facing inwards. The 2.6\,m thick buffer region shields the inner volume against external radioactivity from the PMTs and the SSS. Moreover, the OV serves as a shielding against inward-diffusing radon. The inner components contained inside the SSS are called the \emph{Inner Detector} (ID).  Over time, the number of working PMTs in the ID has decreased. On average, there were 1747, 1576, and 1238 ID PMTs active in Phase-I~\cite{phase1-nusol}, Phase-II~\cite{phase2-nusol}, and Phase-III~\cite{CNO-nature}, respectively. The SSS is enclosed in a cylindrical tank filled with high-purity water, additionally endowed with 208 external PMTs, which define the \emph{Outer Detector} (OD). This water tank serves as an extra shielding against external gammas and neutrons, and as an active Cherenkov veto for residual cosmic muons passing through the detector. There are constant offline checks of the detector's stability and regular online calibrations of PMTs' charge and timing~\cite{phase1-nusol}, to monitor the quality of the acquired data by Borexino.
Borexino detects charged particles that interact with the molecules of the LS based on isotropically emitted scintillation light. The average number of photons produced is typically proportional to the deposited energy and depends on the particle type. However, the energy scale is to some extent intrinsically non-linear due to the ionisation quenching~\cite{Birks} and emission of Cherenkov light~\cite{phase1-nusol}. The PMTs in Borexino then convert the detected light to photoelectrons (p.e.), defined as the electrons removed from the photocathode of the PMT through incident photons. In Borexino, the effective light yield is about 500\,p.e. per 1\,MeV of electron equivalent for 2000 PMTs. This results in $5\%/\sqrt{E~(\text{MeV})}$ energy resolution~\cite{Geo-longpaper}. The Cherenkov light fraction in Borexino is sub-dominant ($<$1\%) and yet fully simulated by the Borexino MC which is typically used for the MC-based spectral fits. A small Cherenkov correction is also considered for the analytical response functions used in Borexino~\cite{phase2-nusol}.

Different energy variables are utilized in Borexino to measure the deposited energy~\cite{Borex-mc, Geo-longpaper}. In this analysis, we use the \emph{Nhits} ($N_{h}^{\mathrm{geo}}$) variable which is defined as the number of photon hits detected by all PMTs. Multiple hits on a single PMT are resolved only if they are more than 180\,ns apart~\cite{phase1-nusol}. The variable is then geometrically normalized, i.e. corrected for two effects: (1)\,since there is a variation in the number of active channels during data-taking, the number of hits is normalized to 2000 channels, and (2)\,since the amount of light seen by a PMT is dependent on its distance to the event, the solid angle of all the PMTs are also taken into account. 

The position reconstruction in Borexino is based on the Time-of-Flight (ToF) technique. The algorithm subtracts from each measured hit time, a position-dependent ToF from the point of particle interaction to the position of the PMT that detected the hit. It then maximizes the likelihood that the event occurs at a certain time and position, based on the measured hit space-time pattern. The maximization
uses the probability density functions of hit
detection, as a function of the time elapsed from the emission of scintillation light~\cite{BxCalibPaper}. The position resolution is about 10\,cm at 1\,MeV at the center of the detector~\cite{Geo-longpaper}, while at larger radii, the resolution decreases on average by a few centimeters. 

Since different particles interact differently in the LS, pulse shape discrimination techniques between $\alpha$ and $\beta/\gamma$ particles and between $\beta^{+}$ and $\beta^{-}$ particles are possible in Borexino~\cite{phase2-nusol, phase1-nusol}. In this analysis, we use the highly efficient Multi-Layer Perceptron (MLP) variable~\cite{Geo-longpaper, CNO-nature}, developed using deep-learning techniques, to distinguish between $\alpha$ and $\beta$ particles in our energy region of interest (Section~\ref{sec:data-sel}).

\section{Solar neutrinos and relevant backgrounds in Borexino}
\label{sec:solar-nu}
\begin{figure}[t!]
    \centering
    \includegraphics[width=0.49\textwidth]{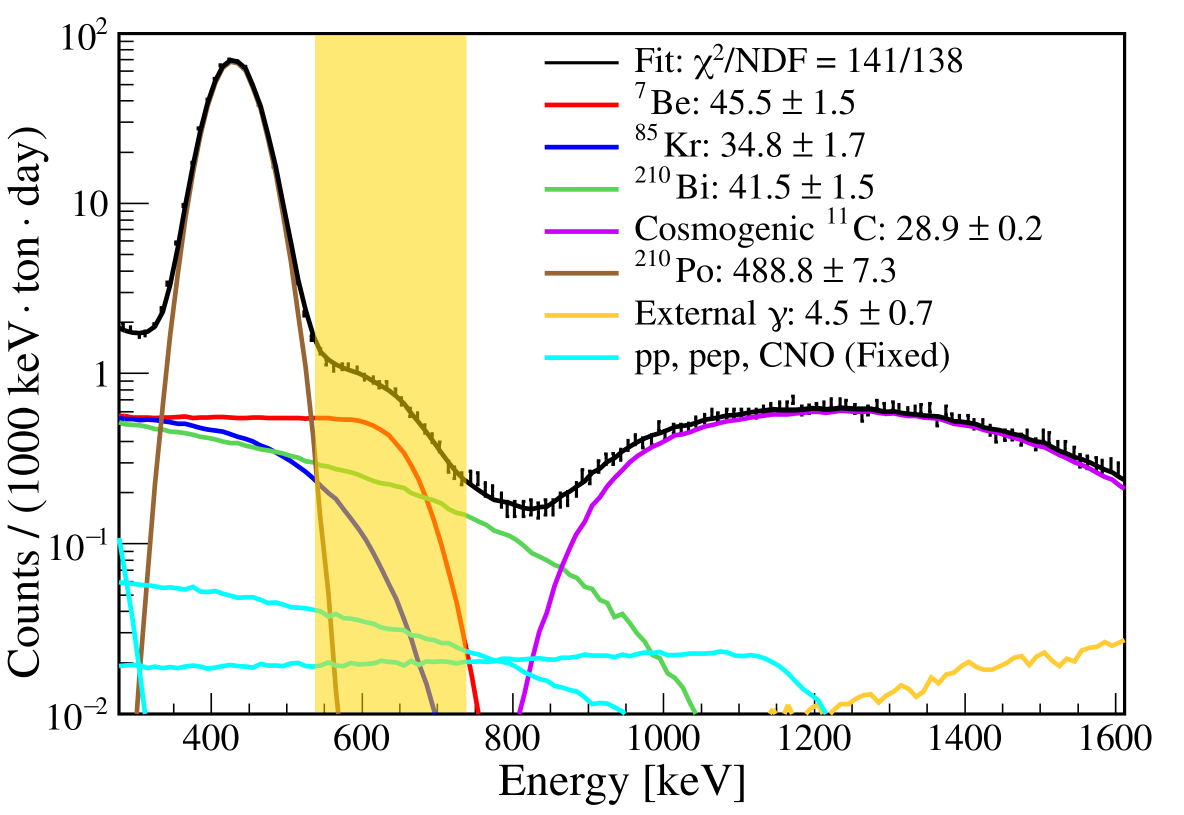}
    \caption{The spectral fit of the energy spectrum performed in Phase-I~\cite{phase1-nusol}. The PDFs of the different solar neutrino components and backgrounds are also shown along with the fitted data. The energy Region of Interest (ROI) used for the directional analysis of Phase-I is shown as a shaded yellow area.}
    \label{fig:nusol-phase2-fit}
\end{figure}
Solar neutrinos are electron-flavor neutrinos produced in the Hydrogen-to-Helium fusion happening inside the Sun via two distinct processes. In the \emph{pp}-chain, responsible for about 99\% of the solar energy, several neutrinos are produced and are named after the particular reaction that produced them. These include mono-energetic neutrinos such as $^{7}$Be\,(0.862\,MeV and 0.384\,MeV) and \emph{pep}\,(1.44\,MeV), as well as neutrinos with continuous energy spectra. The \emph{pp} neutrinos with 0.423\,MeV endpoint dominate the overall solar neutrino flux. The low-flux $^8$B neutrinos extend up to about 16.3\,MeV and are the only neutrinos measured by water Cherenkov detectors such as SuperKamiokande~\cite{super-kamiokande_IV} and SNO~\cite{SNO}. In the sub-dominant CNO-cycle, producing only about 1\% of the solar energy, the fusion is catalyzed by the presence of Carbon, Nitrogen, and Oxygen. The end-point of the CNO solar neutrinos is around 1.74\,MeV.

In Borexino, solar neutrinos are detected via their elastic scattering off electrons. Therefore, even for mono-energetic neutrinos, the spectrum of scattered electrons is continuous, while featuring a Compton-like edge, corresponding to a maximal energy transfer to the scattered electrons. It is impossible to distinguish the recoil electrons of solar neutrinos from the $\beta/\gamma$ background components on an event-by-event basis in Borexino.

In this paper, we analyse events from the energy region at and below the $^{7}$Be Compton-like edge at $\sim$0.66\,MeV (see Section~\ref{sec:data-sel}), where the solar neutrino signal is mostly due to the $^{7}$Be solar neutrinos from the 0.862\,MeV mono-energetic line. The resulting $^{7}$Be rate given in Section~\ref{sec:results} is the sum of both the 0.384\,MeV and 0.862\,MeV mono-energetic lines, which have an interaction rate ratio of 3.6:96.1~\cite{phase1-nusol}. The most precise measurement of the $^{7}$Be neutrinos has been provided by the analysis of the Phase-II dataset~\cite{phase2-nusol,nature-phase2} through a spectral fit in the energy region from 0.19\,MeV to 2.93\,MeV. The Fiducial Volume (FV) used for the spectral fits in Borexino represents the central region of $\sim$70-75\,t, selected to suppress external $\gamma$s from $^{40}$K, $^{214}$Bi, and $^{208}$Tl, originating from the materials surrounding the scintillator. The FV is typically asymmetric and contained within the radius $R<2.8$\,m and the vertical coordinate $-$1.8\,m\,$< z < $\,2.2\,m (Phase-II and III) or $R<3.0$\,m, $-$1.67\,m\,$< z < $\,1.67\,m (Phase-I).

Figure~\ref{fig:nusol-phase2-fit} shows the Phase-I data, along with the various spectral components scaled according to their best fit values obtained via a Poissonian binned likelihood fit. The Region of Interest (ROI) used for the directional analysis of Phase-I in this paper is shown as a shaded yellow band.  

The main backgrounds for the directional analysis include the $\beta$-emitters: $^{85}$Kr($e^{-}$, $Q$=0.687\,MeV) and $^{210}$Bi($e^{-}$, $Q$=1.162\,MeV).  In addition, the distinct peak at about 0.4\,MeV of visible energy is due to $^{210}$Po($\alpha$, $Q$=5.304\,MeV) $\alpha$ background. In the LS, the visible energy of $\alpha$ particles is about an order of magnitude smaller with respect to electrons due to the relatively high ionisation quenching~\cite{Birks}. Therefore, $^{210}$Po is also present in our ROI. However, this background is suppressed heavily in the directional analysis using $\alpha/\beta$ discrimination (Section~\ref{sec:data-sel}).

The data selection criteria of the directional analysis presented in this paper are discussed in Section~\ref{sec:data-sel}. The differences with respect to the standard low energy solar neutrino analysis are a restricted energy interval or ROI between 0.5 and 0.8\,MeV, an enlarged spherical FV, and $\alpha/\beta$ discrimination to suppress $^{210}$Po. In addition, there is a key difference between the standard solar neutrino analysis and the new directional analysis. While the solar neutrino analysis is performed on events, through a spectral fit of their energy distribution, the directional analysis is performed on the properties of individual photon hits detected by the PMTs, for each of the selected events. The principle of this hit-based method is described in the following Section~\ref{sec:cumul_che}.

\section{Correlated and Integrated Directionality (CID)}
\label{sec:cumul_che}

Solar neutrinos interact in the LS via elastic scattering off electrons. 
The angle $\theta_e$ between the recoil electron and the neutrino direction follows the energy-momentum conservation with the free electron approximation:
\begin{equation}\label{eq:scatter_angle}
    \cos{ \theta_e} = \left( 1 + \frac{m_e}{E_\nu} \right)\sqrt{\frac{T}{T+2m_e}},
\end{equation}
where $m_e$ is the electron mass and $T$ is the kinetic energy that the neutrino transfers to the electron, which is deposited in the LS. The LS in turn emits scintillation light with a characteristic wavelength distribution and time profile. At the same time, Cherenkov light is produced in the LS if $T>$~\SI{0.16}{MeV}, as the refractive index of Borexino LS is $\sim${1.55}\,@\,\SI{400}{nm}~\cite{Borex-mc}. Since the Cherenkov spectrum is inversely proportional to the wavelength as $\lambda^{-2}$, most photons produced in the Cherenkov process are in the UV region and thus absorbed and re-emitted as isotropic scintillation light. Only Cherenkov light with a wavelength above the PPO absorption $\lambda>\SI{370}{nm}$~\cite{Borex-mc} retains directional information.
In the chosen energy region for this analysis (Section~\ref{sec:data-sel}), the Monte Carlo gives $\sim$360 Cherenkov photons per event, which corresponds to 1 detected Cherenkov PMT hit per event on average, given the LS absorption, PMT quantum efficiency and detector coverage. At these energies, without the scintillation photons, the events are undetectable just via Cherenkov light due to its scarcity. For these reasons, pure water Cherenkov detectors have not detected sub-MeV solar neutrinos.

\begin{figure*}[t!]
    \centering
    \subfigure[]{\includegraphics[width=0.47\textwidth]{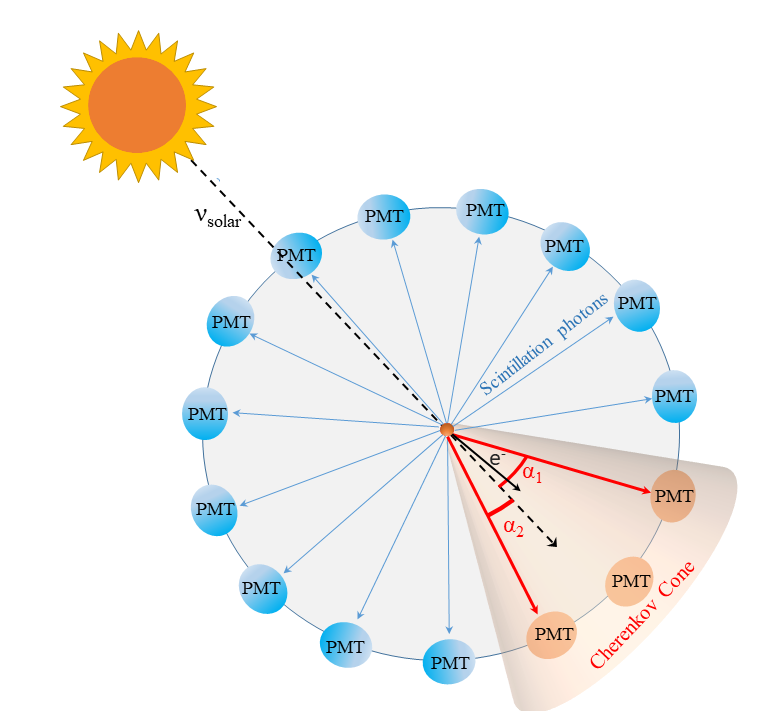}}
    \subfigure[]{\includegraphics[width=0.49\textwidth]{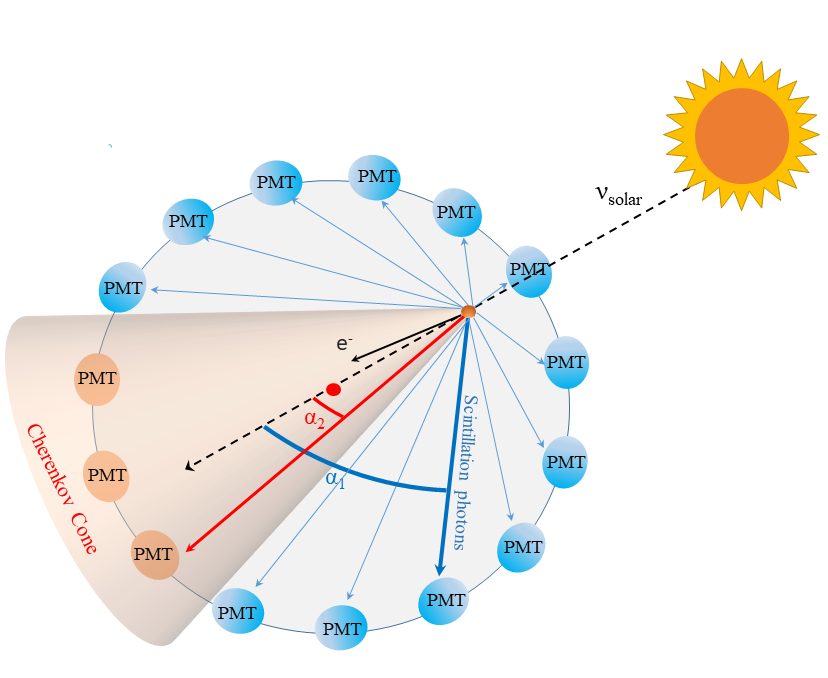}}
    \subfigure[]{\includegraphics[width=0.49\textwidth]{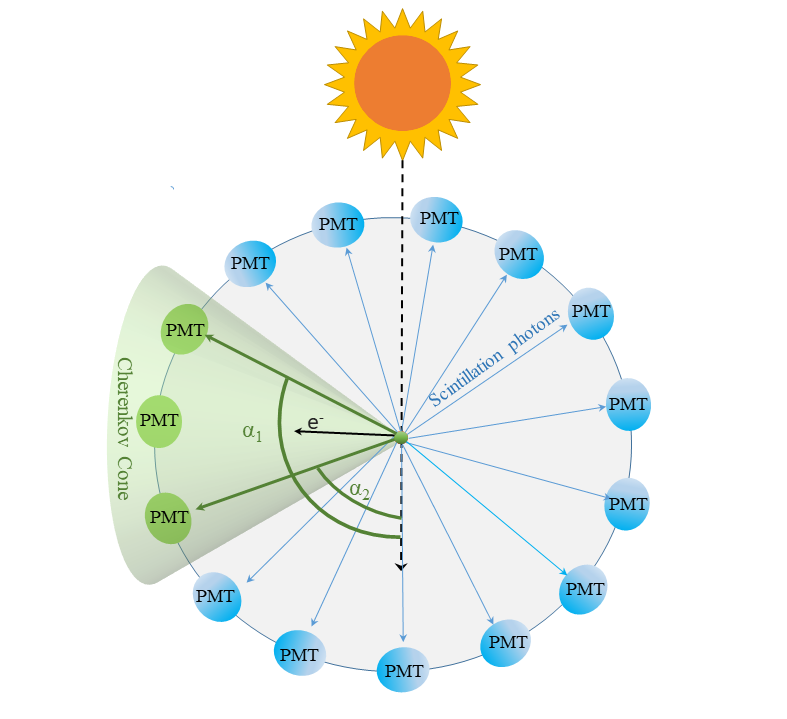}}
    \subfigure[]{\includegraphics[width=0.47\textwidth]{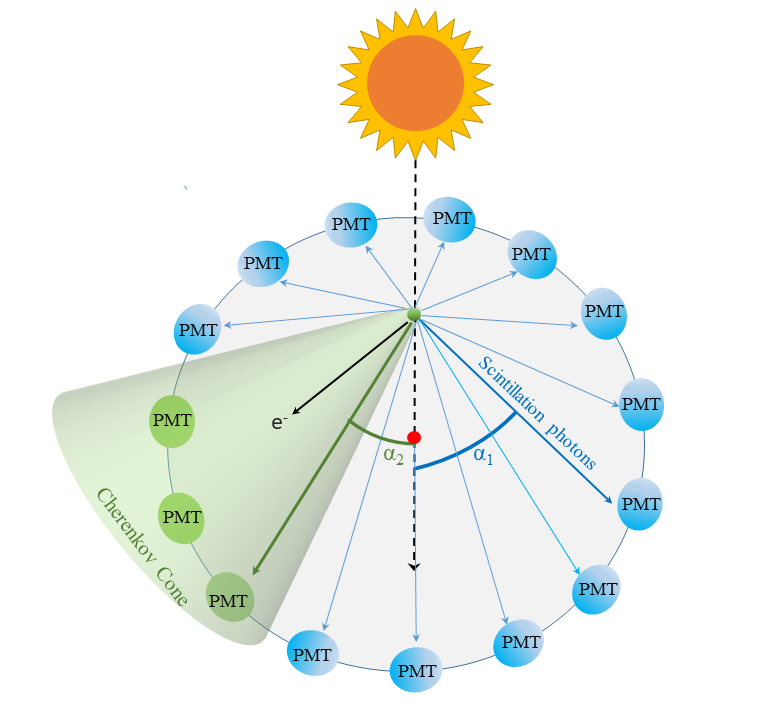}}
    \caption{Schematic representation of the angular correlation, expressed by the angle $\alpha$ between the direction of emitted photons, given by the reconstructed vertex, and the position of the Sun with respect to Borexino for different event types. (a) Electron recoiling off a solar neutrino at the center of the detector produces a Cherenkov cone (red arrows) pointing forward in the direction of the Sun and isotropic scintillation photons (blue arrows). $\alpha_{1}$ and $\alpha_{2}$ are the directional angles of the first and second detected photons of the event, respectively. The Cherenkov photons and, in turn, the PMT hits they trigger are correlated to the incoming direction of the solar neutrinos.
    (b) It is possible that the first detected photon in an event is a scintillation photon, therefore not correlated to the direction of the solar neutrino, and the second detected photon is a Cherenkov photon. Compared to (a), this event results in a flatter angular distribution. In addition, this event happens off-center. (c) An electron from the intrinsic radioactive background also produces a Cherenkov light cone (green arrows) and isotropic scintillation photons (blue arrows). As before, $\alpha_{1}$ and $\alpha_{2}$ are the directional angles of the first and second photons of the background event, respectively. These are Cherenkov photons, but have no correlation to the Sun's direction. (d) Background event similar to (c), but this is an off-center event where the first photon is a scintillation photon and the second detected photon is a Cherenkov photon.}
    \label{fig:cos_alpha_sketch}
\end{figure*}
Figure~\ref{fig:cos_alpha_sketch} shows the principle of the \emph{Correlated and Integrated Directionality (CID)} method: the recoil electron is scattered roughly in the direction of the solar neutrino and the corresponding scintillation and Cherenkov hits are detected at the PMTs. Unabsorbed Cherenkov photons will hit PMTs in the forward direction of the solar neutrino, while scintillation light will hit the PMTs isotropically. The position of the Sun and the direction of solar neutrinos are well-known as events are detected in real time. The $\cos{\alpha}$ can be calculated for each PMT hit, where $\alpha$ is the angle between the known solar neutrino direction and the photon direction of the hit given by the reconstructed position and the hit PMT position. Solar neutrino events and radioactive background in Borexino will produce a PMT hit pattern which is different for scintillation and Cherenkov light (Figure~\ref{fig:cos_alpha_sketch}).
The scintillation light emission is uncorrelated to the solar direction and a flat $\cos{\alpha}$ distribution is expected for both solar neutrinos and background events. The Cherenkov light of solar neutrino events will have a non-flat distribution with a peak at positive $\cos{\alpha}$ values as the recoil electron is forward-scattered. Here, the underlying $\cos{\alpha}$ distribution depends on the energy transfer of the neutrino as in Equation~\eqref{eq:scatter_angle}, as well as the multiple scattering of the recoil electron in the LS.
The Cherenkov light of the radioactive background events is again expected to give a flat $\cos{\alpha}$ distribution as they are not correlated to the solar direction. 
The exact shapes of these $\cos{\alpha}$ CID distributions of signal and background are more complicated due to various effects such as the PMT distribution in the detector (Section~\ref{sec:sys-others}), which can produce deviations from a fully flat $\cos{\alpha}$ distribution.
In principle these hit angle distributions could be analyzed on an event-by-event basis for background rejection. This would require a favorable Cherenkov/scintillation ratio, which is not the case in Borexino. The light yield is much larger for scintillation compared to Cherenkov light, and even at very early times the contribution of the isotropic scintillation light is dominant. In order to achieve some sensitivity, the CID angles of the early PMT hits have to be integrated over a large number of events to produce a $\cos{\alpha}$ distribution.
Given sufficient statistics, a non-flat neutrino signal and a flat background contribution can be fitted to this $\cos{\alpha}$ distribution, making it possible to infer their relative contribution to the total number of events. It should be noted that this summed directionality signal alone is not sufficient to differentiate  between $^{7}$Be, CNO, \emph{pep} solar neutrinos, since these different neutrinos will give indistinguishable CID ($\cos{\alpha}$) distributions  for recoil electrons of similar energy.

\section{Data selection}
\label{sec:data-sel}
\begin{figure}[t!]
    \centering
    \includegraphics[width=0.49\textwidth]{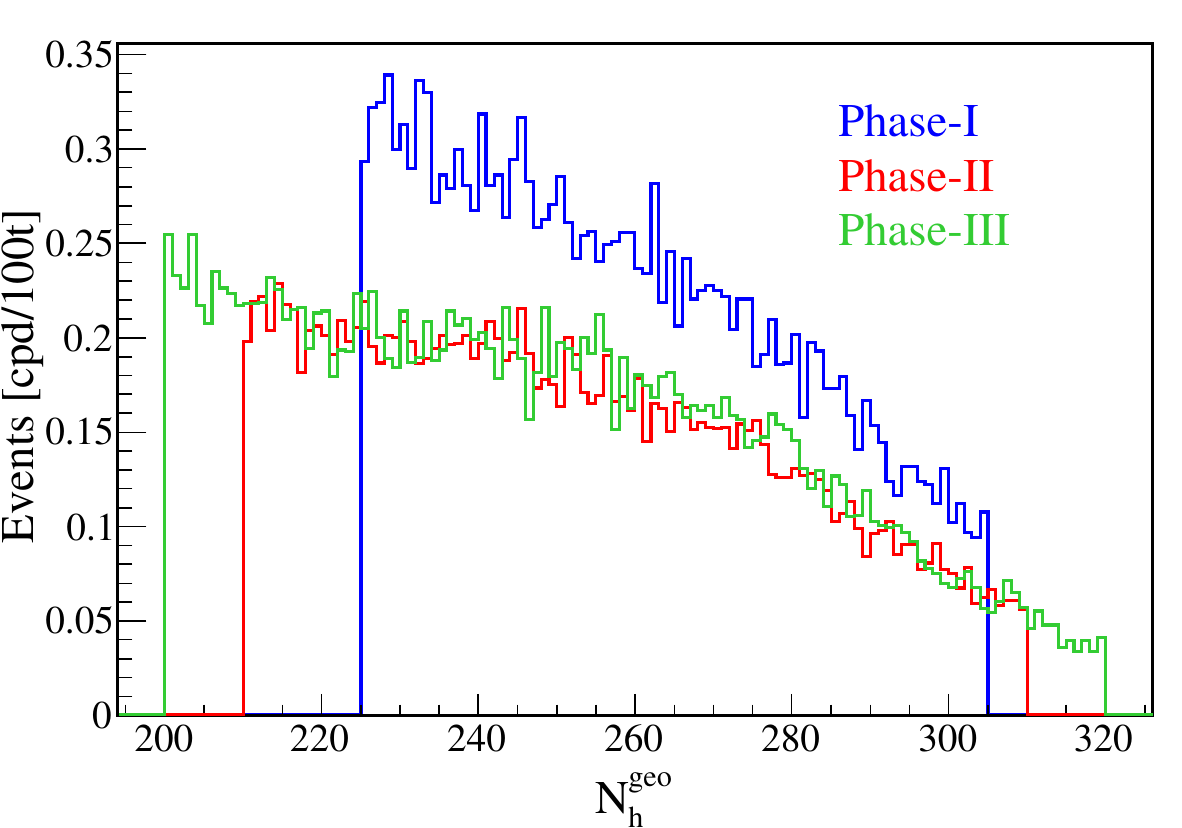}
    \caption{The energy distributions of the data events passing the selection cuts described in Section~\ref{sec:data-sel} in Phase-I (blue), Phase-II (red), and Phase-III (green).}
    \label{fig:energy-data}
\end{figure}
The dataset for this analysis comprises  Phase-I, Phase-II, and Phase-III of the Borexino experiment, corresponding to livetimes  of 740.7~\cite{phase1-nusol}, 1291.5~\cite{phase2-nusol}, and 1072\,days~\cite{CNO-nature}, respectively.

The selection cuts used in this analysis are optimized so as to increase the ratio of $^{7}$Be solar neutrino signal with respect to the radioactive background present in the detector. Most of the applied selection cuts are the same cuts used in the low energy solar neutrino analyses of Borexino~\cite{phase2-nusol}. However, we use an enlarged fiducial volume with a spherical radius of $r$ $<$\,3.3\,m ($<$\,3.0\,m) in Phase-I and II (Phase-III), corresponding to 132.1\,t (99.3\,t), when compared to the fiducial volumes of around 70-75\,t normally used for the spectroscopic analysis (see Section~\ref{sec:solar-nu}).
The bigger fiducial volume has been chosen to increase the statistics of the analysis, since we choose only a specific ROI between 0.5 and 0.8\,MeV in the energy spectrum (Figure~\ref{fig:nusol-phase2-fit}) with a high signal-to-background ratio. This also means that there is no risk of having additional external background ($>$\,1.2\,MeV) which is of concern only for the standard solar neutrino spectroscopy.
\begin{figure}[t!]
    \centering
    \includegraphics[width=0.49\textwidth]{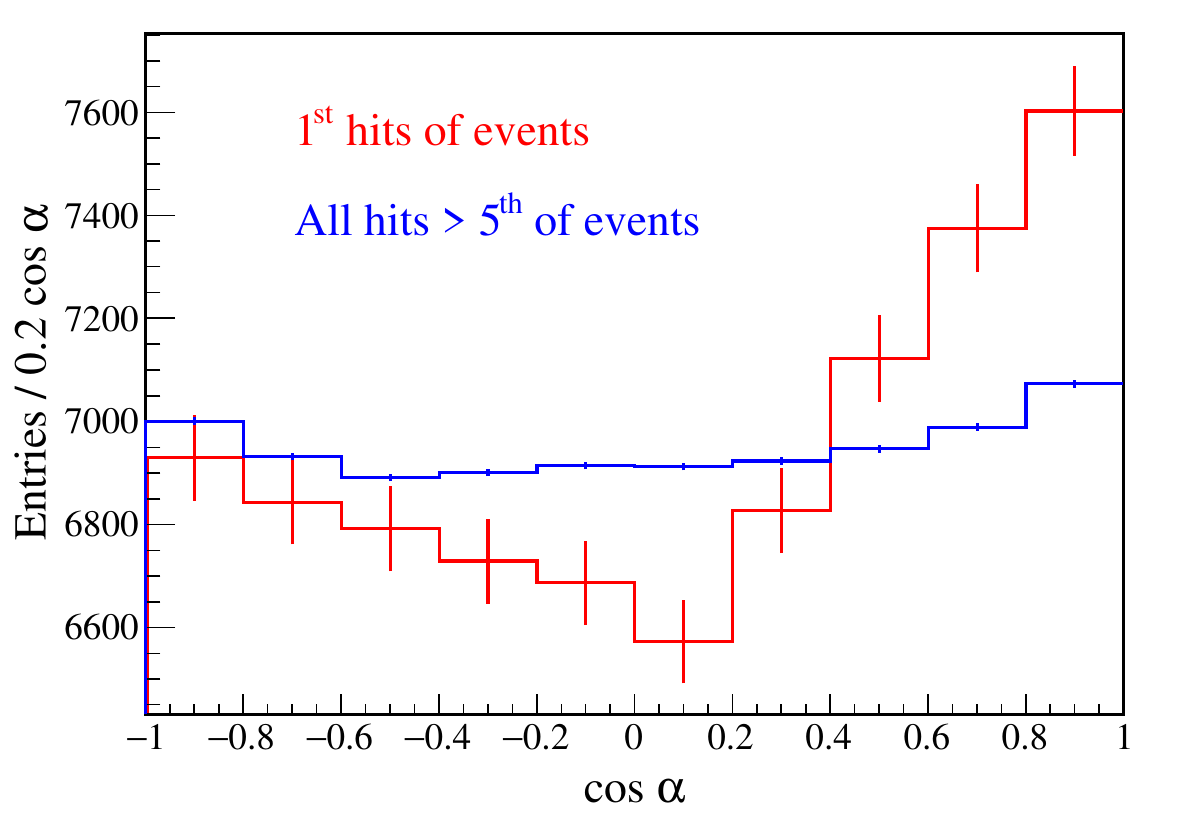}
   \caption{The $\cos{\alpha}$ distribution of the selected data events for Phase-I +  II + III. The very first hit (red) of all the events after Time-of-Flight correction is compared to all the hits detected later than the $>5^\mathrm{th}$ time ordered hit of the events (blue). The histograms are normalized to the same statistics.}
    \label{fig:cosalpha_1-5thhit}
\end{figure}
The signal in the ROI for this analysis consists mainly of $^{7}$Be solar neutrinos ($\sim$90\% of the neutrino signal), and a small amount of \emph{pep} and CNO solar neutrinos. The backgrounds in the ROI include the  $\alpha$-decays from $^{210}$Po, and the $\beta$-decays of $^{210}$Bi and $^{85}$Kr. The energy cuts are applied using the $N_{h}^{\mathrm{geo}}$ variable described in Section~\ref{sec:bxdet}. In Figure~\ref{fig:nusol-phase2-fit}, it can be seen that above the $\sim$0.66\,MeV Compton-like edge of electrons scattered off $^{7}$Be solar neutrinos, there is a steep decrease in the $^{7}$Be rate and the $^{210}$Bi background starts to dominate. Thus, the high energy cut is dependent on the background levels in the different Phases. Since the solar neutrino interaction rate in the detector is homogeneous and not dependent on the radius, only non-homogeneous external backgrounds can increase the event rate in the detector with respect to the radius. Therefore, the high energy cut and the radius of the fiducial volume have been chosen by comparing the event rate in different radii to the standard low energy fiducial volume, in the energy range 0.5-0.8\,MeV.  This method keeps the statistics high, without compromising on the signal-to-background ratio. A high signal ratio also means minimal inclusion of radioactive $^{210}$Po ($\alpha$-background) in the lower energy region $\sim$0.28-0.63\,MeV. For this purpose, we have employed an $\alpha/\beta$ discrimination cut using the MLP variable ($>$0.3) to suppress this background. A Figure of Merit (FOM) estimation is performed using MC PDFs for the optimisation of the lower energy threshold, after fixing the higher energy cut and the radial cut as discussed above. The empirically developed FOM is defined as follows: 
\begin{figure*}[t!]
    \centering
    \subfigure[] {\includegraphics[width=0.49\textwidth]{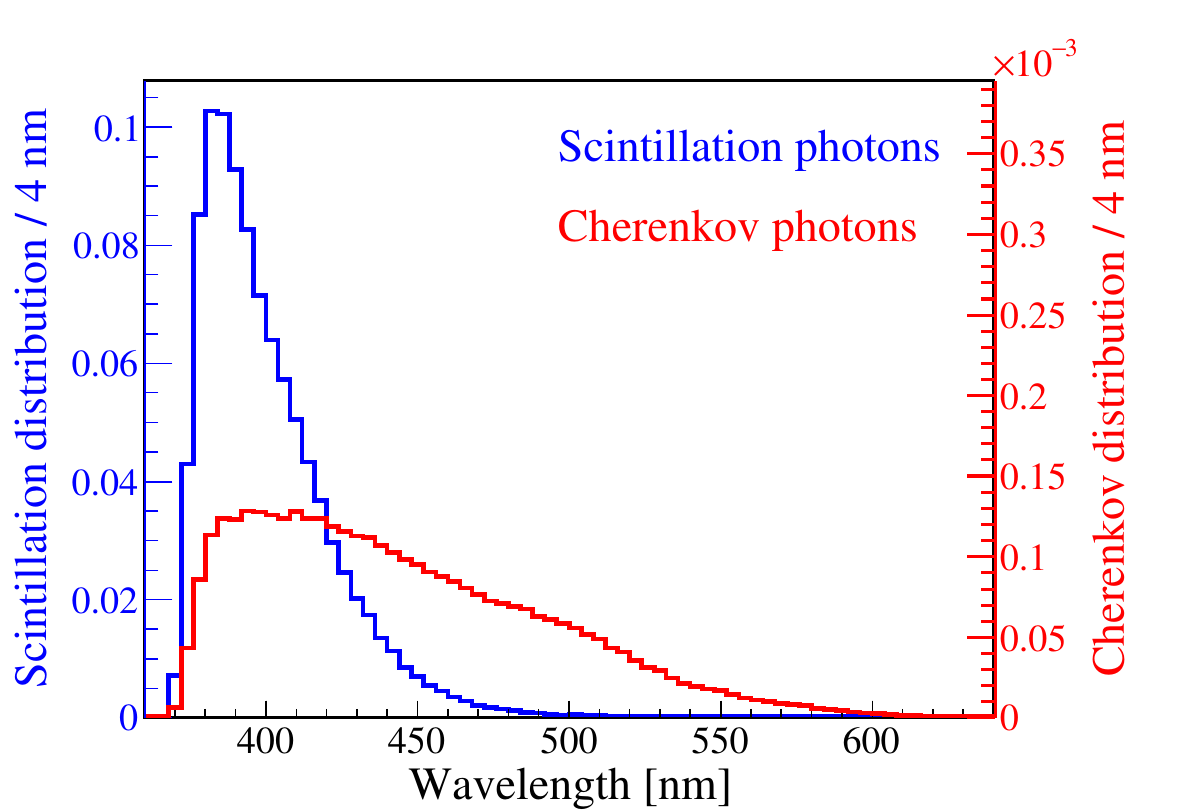}  \label{fig:che_wavelength}}
    \subfigure[]{\includegraphics[width=0.49\textwidth]{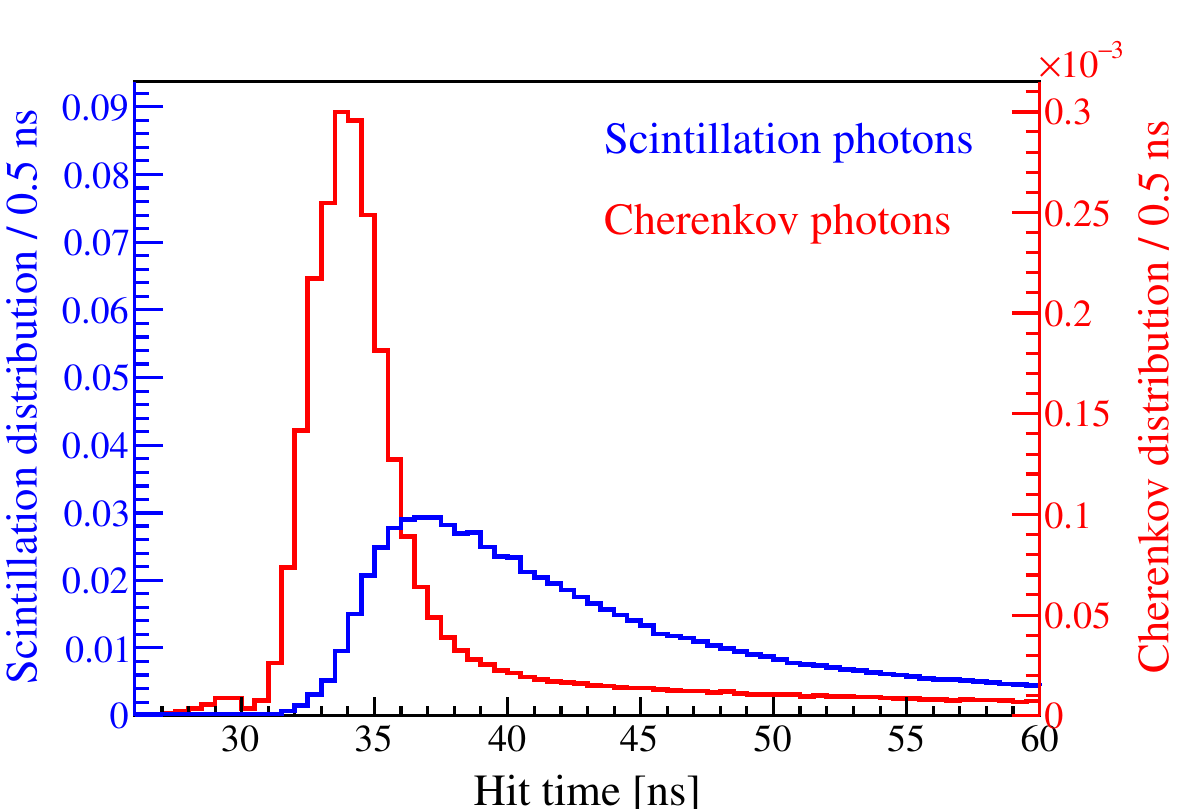}    \label{fig:time-norm}}
    \caption{(a) Borexino Monte-Carlo wavelength spectra as detected by the PMTs. (b) Monte-Carlo distributions of the hit times corrected with their Time-of-Flight for $^{7}$Be solar neutrino recoil electrons (\SI{0.54}{MeV} to \SI{0.74}{MeV}). In both Figures, the left y-axis shows the scintillation (blue), where the area is normalized to 1 and the right y-axis corresponds to Cherenkov light (red), normalized to the number of Cherenkov hits relative to scintillation ($\sim 0.4\%$). Both scintillation profiles also include those photons that have been produced in the Cherenkov process, but have been absorbed and re-emitted by the LS.}
    \label{fig:timepdf}
\end{figure*}
\begin{equation}
    \text{FOM} = \frac{N_{\text{solar}-\nu}}{N_{\alpha-\text{bckg}} + \sqrt{(N_{\text{solar}-\nu} + N_{\beta-\text{bckg}}})},
\end{equation}
where $N_{\text{solar}-\nu}$ is the total number of expected $^{7}$Be, CNO, and \emph{pep} solar neutrino events, $N_{\alpha-\text{bckg}}$ is the number of residual $^{210}$Po events, estimated using the energy-dependent efficiency of the MLP variable, and $N_{\beta-\text{bckg}}$ is the sum of the $\beta$-background components namely, $^{210}$Bi and $^{85}$Kr. The square root in the denominator is used to account for the Poisson statistical errors. However, the residual $^{210}$Po background is taken out of the square root to give more weight to this background and suppress it further. The rates of the solar neutrinos and backgrounds for this estimation have been taken from the results of Borexino's solar neutrino spectroscopy~\cite{phase1-nusol, nature-phase2, CNO-nature}. Since the different Phases have different background levels, the optimisation results in different ROIs of 225\,$<N_h<$\,305, 210\,$<N_h<$\,310, and 200\,$<N_{h}<$\,320, in Phase-I, II, and III, respectively.

The energy spectra of the selected data events for the three Phases are shown in Figure~\ref{fig:energy-data}. The CID $\cos{\alpha}$ distribution of the selected data events in all three phases is shown in Figure~\ref{fig:cosalpha_1-5thhit}. The first hit (red) after Time-of-Flight correction of all the events is compared to the later hits $>5^{\text{th}}$ (blue) of the events. The peak at $\cos{\alpha}>$0 is clearly visible for the first hit as the Cherenkov light contributes significantly only at the earliest hits of an event. For the sum of all the later hits there is no directional signature visible since the isotropic scintillation light is dominant. This distribution is not perfectly flat due to the distribution of live PMTs and other effects, explained in more detail in Section~\ref{sec:sys-others}. The number of hits to be used for the analysis has been optimized using MC and is discussed in Section~\ref{sec:analysis}.

\section{Cherenkov and scintillation light in Borexino}
\label{sec:mc_prod}

Borexino uses a customized Geant-4 based Monte-Carlo code which simulates the entire detection process following a particle interaction in the detector~\cite{Borex-mc, GEANT4}. The parameters of the MC simulation have been tuned with the use of calibration data acquired with radioactive sources inserted in the Borexino detector~\cite{BxCalibPaper}. For Cherenkov light there has been no dedicated calibration, as it is expected that unabsorbed Cherenkov light only has a negligible influence on the position reconstruction and other algorithms used in Borexino. So, the precise time profile of Cherenkov light is of no interest for the spectroscopic Borexino analyses.

The Cherenkov photons necessary to perform the CID analysis are simulated according to the Frank–Tamm formula~\cite{particle_data_group}. Cherenkov photons are produced in the LS if the kinetic energy of the electron $T$ is $>$\SI{0.16}{MeV}, as the refractive index of Borexino LS is $\sim${1.55}\,@\,\SI{400}{nm}~\cite{Borex-mc}. Their spectrum as well as their velocity in the scintillator depends on the wavelength-dependent refractive index n($\lambda$) (see Figure\,9 in~\cite{Borex-mc}), which is given to the MC from a laboratory measurement with an uncertainty of $\sim$1\%\,\cite{Borex-mc}. The wavelength spectrum detected by the PMTs according to MC can be seen in Figure~\ref{fig:che_wavelength}. Both scintillation (blue) and Cherenkov (red) spectra start sharply at \SI{370}{nm}. This is due to the absorption of the light in the LS and its re-emission by the PPO~\cite{Borex-mc}. All Cherenkov light with a wavelength $\lambda<\SI{370}{nm}$ is absorbed and re-emitted as scintillation light.

The Cherenkov light is emitted instantly while the scintillation light emission follows a multi-exponential decay time where the fastest component has \SI{1.6}{ns}~\cite{Borex-mc}. This intrinsic time distribution is then smeared by various optical processes during the light propagation, Transit Time Spread of the PMTs, as well as the precision of the PMT time calibration. Figure~\ref{fig:time-norm} shows the Time-of-Flight corrected PMT hit time distribution of Cherenkov and scintillation photons after the full simulation and reconstruction chain. The time axis is normalized such that the beginning of the scintillation time profile corresponds to the ToF of photons from the center of the detector. The distributions are normalized to their area to show the relative time behavior. However, according to the MC, the ratio between all Cherenkov and  scintillation hits in the energy ROI is $\sim$0.4\%. Therefore, some form of a time cut is necessary to increase the directional sensitivity, as it will be explained in Section~\ref{sec:analysis}. The relative time distribution for Cherenkov and scintillation photons can be different between Data and MC, which has an influence on the CID analysis. For this, a calibration of the effective Cherenkov group velocity has been performed and will be discussed in Section~\ref{sec:gamma_calib}.

\section{Analysis strategy and methods}
\label{sec:analysis}

The main goal of this analysis is to show that using the \emph{Correlated and Integrated Directionality (CID)} method it is possible to provide a statistically significant measurement of the number of solar neutrinos ($^{7}$Be, \emph{pep}, CNO) in the favorable energy region featuring the $^{7}$Be solar neutrinos. The CID method works by correlating the very first hits of the recoil electrons from solar neutrinos to the known position of the Sun. The number of solar neutrinos $N_{\mathrm{solar}-\nu}$ can be measured by performing a $\chi^{2}$-fit on the CID ($\cos{\alpha}$) distribution of data, using the MC directional PDFs of $^{7}$Be and $^{210}$Bi. The measurement can then be converted to the $^{7}$Be neutrino interaction rate, expressed in counts per day (cpd) per 100\,ton of  LS, using the known exposure and the efficiency of the selection cuts, and the Standard Solar Model (SSM) predictions of the CNO and \emph{pep} neutrino rates.
\begin{figure*}[t!]
    \centering
    \subfigure[]{\includegraphics[width=0.49\textwidth]{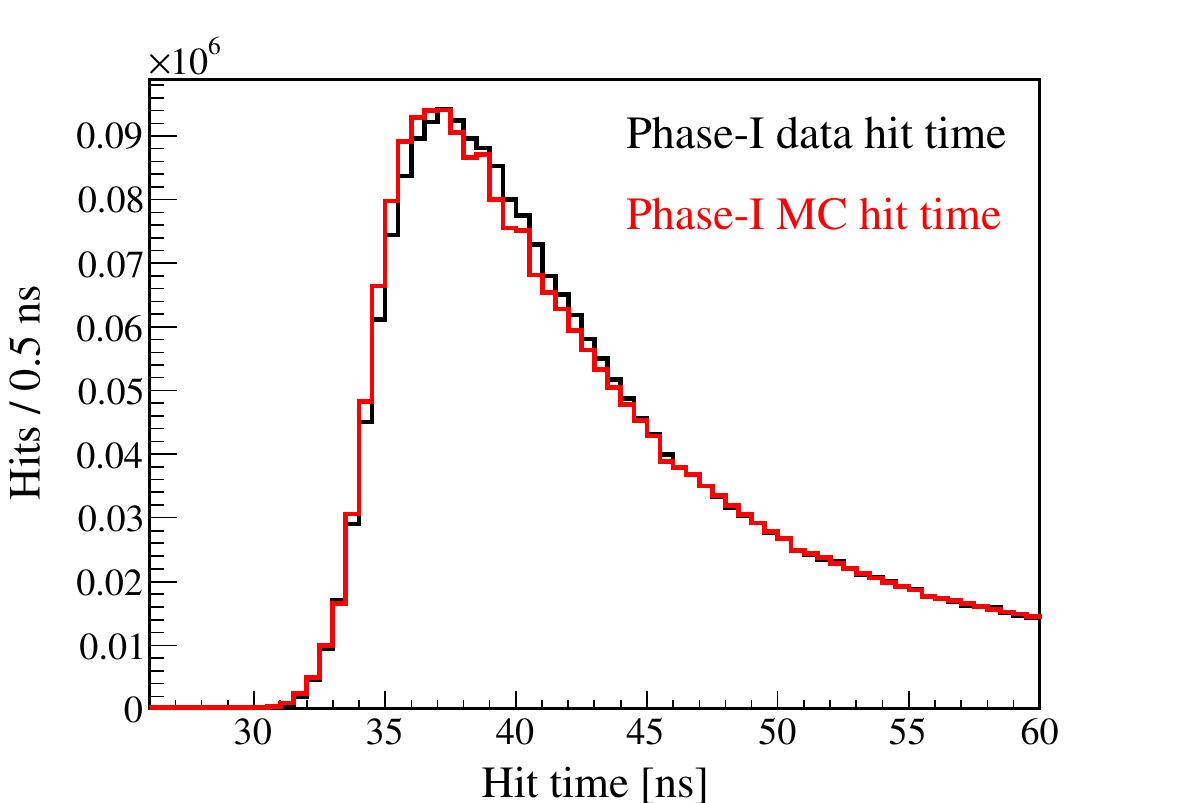}    \label{fig:hittime_data_mc}}
    \subfigure[]{\includegraphics[width=0.49\textwidth]{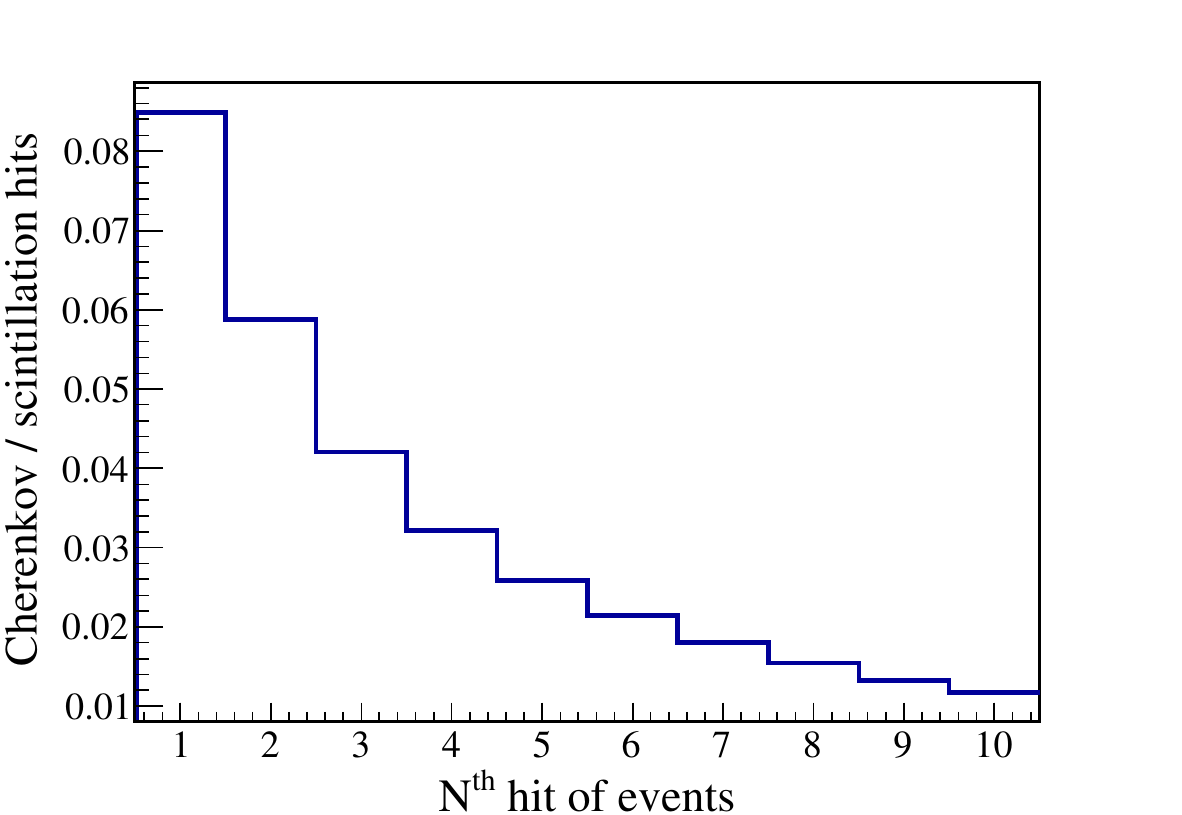}   \label{fig:nthhit_ratio_che_scint}}
    \caption{(a) Time-of-Flight corrected hit times for Phase-I data (black) and MC (red) in the ROI, where the MC is normalized to data statistics. Both $^{7}$Be and $^{210}$Bi MC are comparable within their statistics and only $^{7}$Be is shown. (b) Ratio of detected Cherenkov-to-scintillation hits as a function of the Time-of-Flight sorted hits for all events of $^{7}$Be MC.}
\end{figure*}

The Borexino MC is used to produce the CID PDFs with the expected $\cos{\alpha}$ angle distributions for the solar neutrino signal and for the radioactive background. The recoil electrons of $^{7}$Be solar neutrinos are simulated according to their theoretical energy spectrum and the angle relative to the neutrino direction following Equation~\eqref{eq:scatter_angle}. In the chosen ROI, the average $\theta_e$ between recoil electrons and $^{7}$Be solar neutrinos is $\sim$16$^{\circ}$. The Borexino MC simulates the entire detection chain, including multiple scattering of electrons in the LS, electronics simulation, and event reconstruction. The $^{7}$Be solar neutrinos are the dominant contribution to the signal. Other solar neutrinos such as CNO and \emph{pep} neutrinos, contribute at the level of 6-7\% to the total event rate (see Section~\ref{sec:data-sel}). Although the angle between the scattered electron and the neutrino is dependent on the electron and neutrino energies (Equation~\eqref{eq:scatter_angle}), the difference between the $\cos{\alpha}$ distributions of CNO, \emph{pep}, and $^{7}$Be solar neutrinos in our ROI have been found to be negligible. For the background, we simulate the dominant $^{210}$Bi (see Figure~\ref{fig:nusol-phase2-fit}) and all other backgrounds have the same $\cos{\alpha}$ distribution, as they are isotropic and uncorrelated to the position of the Sun. The $^{210}$Po background is already well-suppressed due to the $\alpha/\beta$ discrimination cut and the FOM optimization, as described in Section~\ref{sec:data-sel}. The simulation of this $^{7}$Be signal and $^{210}$Bi background is performed on an event-by-event basis. For each individual data event, that corresponds to a certain position of the Sun, 200 MC events are simulated within a sphere of \SI{15}{cm} radius around the reconstructed position of the data event and with the set of PMTs that were active in that particular moment. Thus the possible hit patterns of the active PMTs in relation to the position of the Sun and the event position are correctly taken into account in the MC production. This procedure is repeated for each event twice\textemdash once assuming it is the $^{7}$Be signal and once assuming it is the $^{210}$Bi background, according to their spectral shape within the energy region contributing to the ROI. This is an intrinsically different approach to the previous Borexino analyses, where the events are simulated uniformly in the whole FV for the analyzed period. The same data selection cuts described in Section~\ref{sec:data-sel} are then applied on the MC simulated events as well.

An event-by-event directional reconstruction is not possible in Borexino as discussed in Section~\ref{sec:cumul_che}, and therefore, we study the superimposed PMT hit distributions of all data events. For each event, these hits are then sorted in time ($t$) with respect to the reconstructed start time of the event ($t_{0}$), corrected with their Time-of-Flight (ToF), i.e. $t-t_{0}-\mathrm{ToF}$ in both data and MC. The start time of the event $t_{0}$ is calculated such that the beginning of the scintillation time profile corresponds to the ToF of photons from the center of the detector. The ToF correction guarantees that the time distribution is comparable for events happening in the center and the edge of the detector. The ToF is calculated as $\text{ToF} = n_{\text{eff}}\cdot d_{\text{PMT}}/c$, where $n_\text{{eff}}$ is the effective refractive index of the Borexino LS, $d_{\text{PMT}}$ is the distance between the reconstructed event position and the PMT that detected the photon hit, and $c$ is the speed of light in vacuum. The effective refractive index $n_\text{{eff}}$ represents the group velocity for the detectable wavelength distribution of the Borexino LS. It has been measured as (1.6631\,$\pm$\,0.0005) for data and (1.6531\,$\pm$\,0.0003) for MC. These $n_\text{{eff}}$ values are based on an improved estimation performed for this analysis using the $^{14}$C-$^{222}$Rn calibration source~\cite{BxCalibPaper}, specifically on the $^{214}$Po $\alpha$ decays ($Q$=6.0\,MeV) from the $^{222}$Rn chain. These values are slightly different from the initial estimates (1.68 for data and 1.66 for MC~\cite{Borex-mc}), tuned on the position reconstruction algorithm. However, the results are compatible within the instrumental error on the measurement of the refractive index~\cite{Borex-mc}. The difference between the data and MC values (0.6\%) is also consistent with this instrumental error. 

Systematic sub-nanosecond time differences have been observed between different PMTs through a study performed using the same $^{14}$C-$^{222}$Rn calibration source. The estimated time corrections have been applied on the time of the photon hits from different PMTs in data before sorting them. These systematic differences between PMTs are relevant only for the directionality analysis. Since the $\cos{\alpha}$ distribution also depends on the position of the PMTs in the detector, only hits from PMTs whose behavior are correctly reproduced in the MC simulation are selected. The systematic uncertainty due to the method of selection of the PMTs is discussed in Section~\ref{sec:sys-others}.
\begin{figure*}[t!]
    \centering
    \subfigure[]{\includegraphics[width=0.47\textwidth]{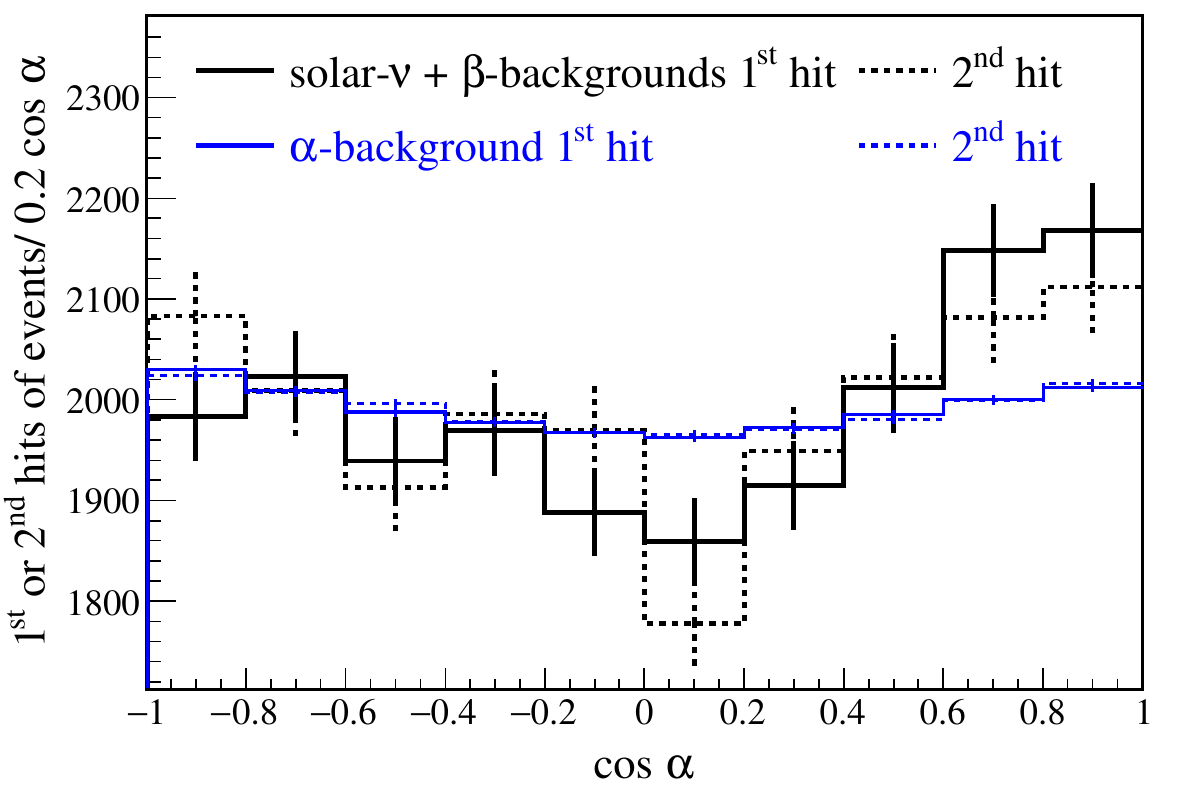}  \label{fig:cos_alpha_data}} 
    \subfigure[]{\includegraphics[width=0.51\textwidth]{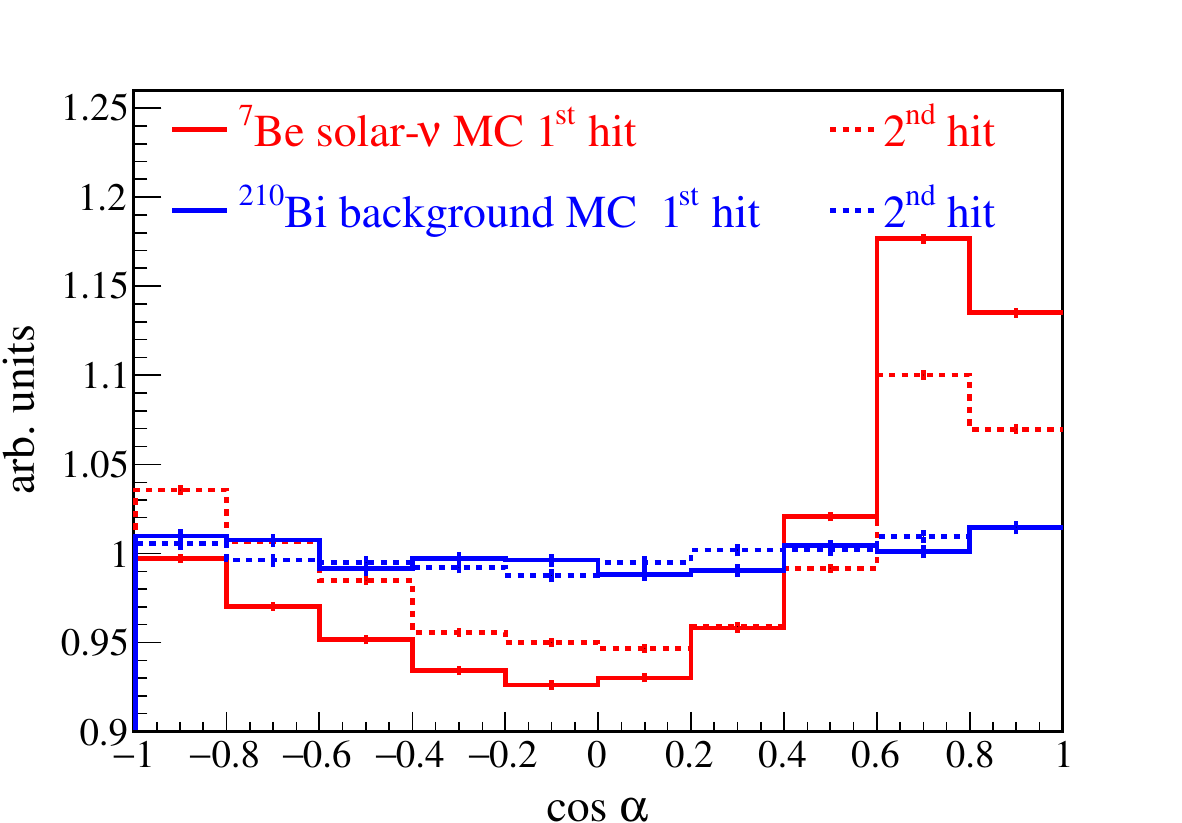}  
    \label{fig:cos_alpha_mc}} 
    \caption{Distributions of the $\cos{\alpha}$ directional angle. The first hits are shown as solid lines and the second hits are shown as dotted lines.
    (a) $\cos{\alpha}$ distribution of data events from Borexino Phase-I, i.e. 19904 $\beta$-events consisting of solar neutrinos + $\beta$-backgrounds (black) chosen using the selection cuts as in Section~\ref{sec:data-sel} and 1.8\,million $^{210}$Po ($\alpha$) background events (blue) in Phase-I. The histograms are normalized to have 19904 entries. A clear peak at positive $\cos{\alpha}$ values can be observed only for the solar neutrino-rich sample. (b) $\cos{\alpha}$ distributions for MC events of $^{7}$Be solar neutrinos (red) and $^{210}$Bi background (blue). The histograms are normalized to have the same area. The first hit (solid line) of $^{7}$Be MC carries more directional information than the second hit (dotted line).}
\end{figure*}

A straightforward approach to select Cherenkov photons would be to apply a time cut to maximize the Cherenkov to scintillation ratio according to MC. However, small systematic differences have been observed between the hit time distributions of data and MC, particularly for the early times rich in Cherenkov photons. This can be seen in Figure~\ref{fig:hittime_data_mc}, where data and MC are significantly different for the large statistics of the summed data hits. Within the statistics of a single event these time distributions are comparable and therefore produce comparable results for event-based algorithms such as the position reconstruction. The observed differences make it challenging to select Cherenkov photons by applying a cut on the absolute time. Despite the time differences, the \emph{relative} ordering of the photon hits is still in agreement between MC and data. Consequently, we adopt an {\it``N\textsuperscript{th}-hit method''} to select single photon hits of an event, ordered in time after the ToF subtraction. First, we construct $\cos{\alpha}$ distributions of the 1\textsuperscript{st},2\textsuperscript{nd},...,N\textsuperscript{th} hits of all the selected events. Then, in order to maximize the amount of Cherenkov photons, a cut is applied on the N\textsuperscript{th} hit. This N\textsuperscript{th} hit cut is chosen based on the $\chi^{2}$ between the $\cos{\alpha}$ MC PDFs of $^{7}$Be and $^{210}$Bi. It has been found that the first two hits of all the events are the most sensitive to the directional differences between the signal and the background MC. Figure~\ref{fig:nthhit_ratio_che_scint} shows the ratio of Cherenkov to scintillation light with respect to the N\textsuperscript{th} hit, where a steady decrease can be seen for later hits. While the first hits of all events have the largest amount of directional information due to the significant contribution of Cherenkov light at the earliest times (Figure~\ref{fig:timepdf}), some of the information is carried also by the second hit. Although later hits have a Cherenkov contribution, they carry less directional information, as these Cherenkov photons are more likely to have been affected by effects such as scattering.

Figure~\ref{fig:cos_alpha_data} shows the $\cos{\alpha}$ distributions of the first two hits of $\alpha$-like (blue) and $\beta$-like (black) events in Phase-I. The $\beta$-events (19904) are chosen using the selection cuts of this analysis (Section~\ref{sec:data-sel}) and include solar neutrinos and other $\beta$-backgrounds. The $\alpha$ events (1.8\,million) consist of $^{210}$Po background. It can already be seen from these data that there is a clear peak at $\cos{\alpha}\,>$0.6 for the solar neutrino-rich sample, when compared to the background-only $\alpha$ events that have no correlation to the Sun's position. The peak at negative $\cos{\alpha}$ is due to a small bias between the true and reconstructed position of the electron events as it will be described later in this Section. The second hits (black dotted) show a similar shape to the first hits (black solid), but the peak at $\cos{\alpha}\,>$0.6 is less pronounced, which is expected as there is a smaller Cherenkov photon ratio compared to the first hits. Figure~\ref{fig:cos_alpha_mc} shows the $\cos{\alpha}$ distribution of the first two hits of $^{7}$Be (red) and $^{210}$Bi (blue) MC simulated events, normalized to have the same area. The simulated $^{210}$Bi background (blue solid and dotted lines) shows a flatter $\cos{\alpha}$ distribution for both the first and second hits. However the distributions are not perfectly flat, and this is due to multiple effects studied individually, as described in Section~\ref{sec:sys-others}.
\begin{figure*}[t!]
   \centering
    \subfigure[]{\includegraphics[width=0.4\textwidth]{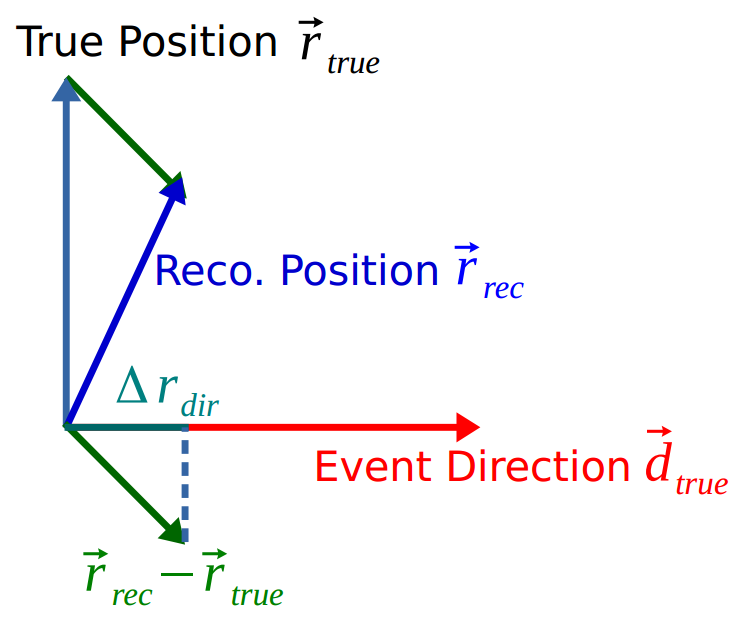}    \label{fig:pos_reco_sketch}}
    \subfigure[]{\includegraphics[width=0.5\textwidth]{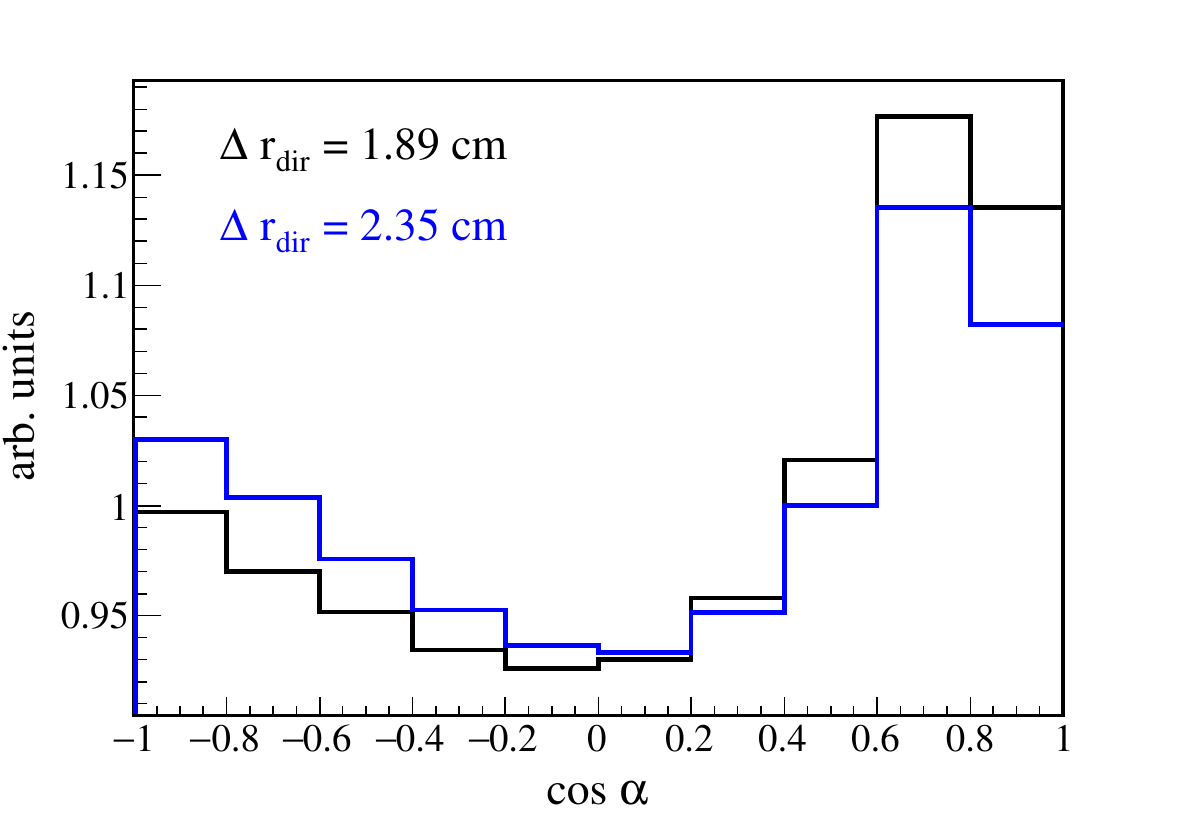}    \label{fig:pos_reco_cosalpha}}
    \caption{(a) Effect of the biased position reconstruction $\Delta r_{\mathrm{dir}}$ observed in MC where the $e^{-}$ position is reconstructed slightly towards the direction of the $e^{-}$. This bias of $\sim$\SI{2}{\centi\meter} is not visible on an event-by-event basis due to the position reconstruction resolution of \SI{12}{\centi\meter} in the ROI. (b) The effect of this parameter, where the normal $^{7}$Be MC has $\Delta r_{\text{dir}}$=1.89\,cm (black). For comparison, a larger $\Delta r_{\text{dir}}$=2.35\,cm (blue) increases the slope at $\cos{\alpha}$=-1.}
\end{figure*}

The number of solar neutrinos $N_{\mathrm{solar}-\nu}$ for the selected $\beta$-like data events (as in Figure~\ref{fig:cos_alpha_data}) is extracted from the fit of their $\cos{\alpha}$ distribution with the MC based PDFs, i.e., $\cos{\alpha}$ distributions of $^7$Be signal and $^{210}$Bi background (as in Figure~\ref{fig:cos_alpha_mc}), using a $\chi^{2}$-fit defined as follows: 
\begin{align}\label{eq:nusol-fit}
\nonumber
    &\chi^{2}(N_{\mathrm{solar}-\nu}) = \\ \nonumber
    & \sum \limits_{n=1}^{N}\sum\limits_{i=1}^{I}
    \left(
    \frac{\bigg((\cos{\alpha})^{D}_{n, i} - (\cos{\alpha})^{M}_{n, i}\big(N_{\mathrm{solar}-\nu}, \Delta r_{\mathrm{dir}}, gv_{\mathrm{ch}}^{\mathrm{corr}}\big)\bigg)^{2}}{(\sigma^{D}_{n, i})^{2} + (\sigma^{M}_{n, i})^{2}} 
    \right.\\ 
    &+ \left.\frac{(gv_{\mathrm{ch}}^{\mathrm{corr}} - \SI{0.108}{\nano\second\per\meter})^2}{(\SI{0.039}{\nano\second\per\meter})^2} 
    \right)
\end{align}
The index $n$ runs from 1 to the selected N\textsuperscript{th} hit $N=2$ and the index $i$ runs from 1 to the total number of bins $I$ in the range -1\,$< \cos{\alpha} <$\,+1. $(\cos{\alpha})^{D}_{n, i}$ and $(\cos{\alpha})^{M}_{n, i}$ are the $\cos{\alpha}$ values for the i\textsuperscript{th} bin of the n\textsuperscript{th} hit of data and MC, respectively, and, $\sigma^{D}_{n,i}$ and $\sigma^{M}_{n, i}$ are their corresponding statistical errors. The systematic shift in the reconstructed vertex position of the electron $\Delta r_{\mathrm{dir}}$ and an effective correction of the group velocity $gv_{\text{ch}}^{\mathrm{corr}}$ for Cherenkov photons are sufficient to parameterize  the differences between data and MC. The group velocity correction $gv_{\text{ch}}^{\mathrm{corr}}$ is applied for Cherenkov photons in the MC, estimated using gamma calibration sources and it is treated as a nuisance parameter using a Gaussian pull term for Phase-I. This effective correction is discussed in detail in Section~\ref{sec:gamma_calib} and the best fit value has been estimated as (0.108 $\pm$ 0.039)\,ns\,m$^{-1}$. The systematic parameter $\Delta r_{\mathrm{dir}}$, arises due to the difference between the true and reconstructed positions of the detected electron, and is treated as a free nuisance parameter in the fit and is explained below.

In this analysis the direction of the photon hits are calculated based on the reconstructed position of the neutrino recoil electron in the LS and the PMT position that detected the hit. This means that the directions of the hits depend on the accuracy of the reconstructed position of the scattered electron. Using the known true MC position $\vec{r}_{\text{true}}$, as well as the true direction of the simulated recoil electron $\vec{d}_{\text{true}}$, the bias of the reconstructed position $\vec{r}_{\text{rec}}$ is given by the mean of $\Delta r_{\text{dir}}$ of many MC recoil electron events:
\begin{equation}\label{eq:delta_r_dir_def}
    \Delta r_{\text{dir}} = \left(\vec{r}_{\text{rec}} - \vec{r}_{\text{true}}\right)\cdot \vec{d}_{\text{true}}.
\end{equation}
This equation is schematically represented in Figure~\ref{fig:pos_reco_sketch}. The effect of this biased position reconstruction in the $^{7}$Be CID MC is shown in Figure~\ref{fig:pos_reco_cosalpha}, where a larger $\Delta r_{\text{dir}}$ shows a bigger negative slope at $\cos{\alpha}<$1 for the same number of Cherenkov photons in both histograms. The expected value for $\Delta r_{\text{dir}}$ in MC is \SI{1.89}{cm} over a position resolution of \SI{12}{\centi\meter} in the ROI. Relatively small changes in $\Delta r_{\text{dir}}$ have a large impact on the $\cos{\alpha}$ shape. This effect is not present in background, in which the true electron direction is not correlated to the position of the Sun. The value of this effect in data is unknown due to the lack of a dedicated $e^-$ Cherenkov calibration in Borexino. Thus, $\Delta r_{\text{dir}}$ is left as a free nuisance parameter in the final fit presented in Section~\ref{sec:results} and is allowed to vary.

The binning for the $\cos{\alpha}$ distribution has been chosen based on a MC study. For a fixed injected $N_{\mathrm{solar}-\nu}$, the standard deviation of the distribution of the extracted $N_{\mathrm{solar}-\nu}$ showed a steady decrease until 20 bins, followed by a stable precision between 20 and 80 bins. Therefore, 60 bins has been chosen for the $\cos{\alpha}$ histograms used in the final fit. The systematic uncertainty due to the choice of binning is discussed in Section~\ref{sec:sys-others}.

The \emph{``N\textsuperscript{th}-hit method''} has also been validated using a MC study. The difference between the time distributions of data and MC can be empirically described with a Gaussian derivative. In the MC study, this difference has been fitted with such a function and has been used to change the MC time distribution so as to make it similar to data. From this modified MC time distribution, pseudo-datasets with a fixed amount of $N_{\mathrm{solar}-\nu}$ have been then constructed. The normal MC PDFs without any corrections have then been used to perform the $\chi^{2}$ fit. A 13\% bias has been observed between the injected and extracted $N_{\mathrm{solar}-\nu}$, for a simple time cut ($<$ 34\,ns), while no bias has been found for the ``N\textsuperscript{th} hit method''. This indicates that the method chosen for the CID analysis, makes it possible to suppress the absolute systematic time differences between data and MC. 

\section{Gamma Cherenkov calibration}
\label{sec:gamma_calib}
\begin{figure}[t]
    \centering
    \includegraphics[width=0.5\textwidth]{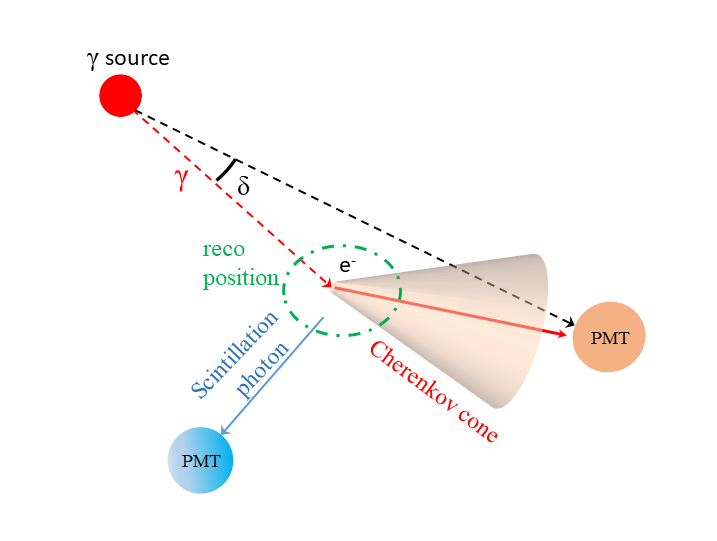}
    \caption{Schematic representation of the $\cos{\delta}$ angle used for the Cherenkov calibration with gamma sources. A $\gamma$ (red dashed line) is emitted from the calibration source position $\vec{r}_{\mathrm{source}}$ (red filled circle). The Compton-scattered $e^-$s (red solid line) in turn emit Cherenkov light (orange cone) in correlation to the direction of the incident $\gamma$. The direction of scintillation photons (blue) are uncorrelated to the $\gamma$ direction.
    The directional angle $\delta$ used for the calibration of the Cherenkov light in this analysis is the angle between the reconstructed gamma direction (based on the reconstructed position $\vec{r}_{\mathrm{rec}}$ (green circle)) and the photon direction of the hit (based on the known source position and the PMT position $\vec{r}^{\hspace{0.5mm}\mathrm{PMT}}$), as defined in Equation~\eqref{eq:cosdelta_def}.}
    \label{fig:cos_delta_sketch}
\end{figure}
\begin{figure*}[t!]
    \centering
    \subfigure[]{\includegraphics[width=0.49\textwidth]{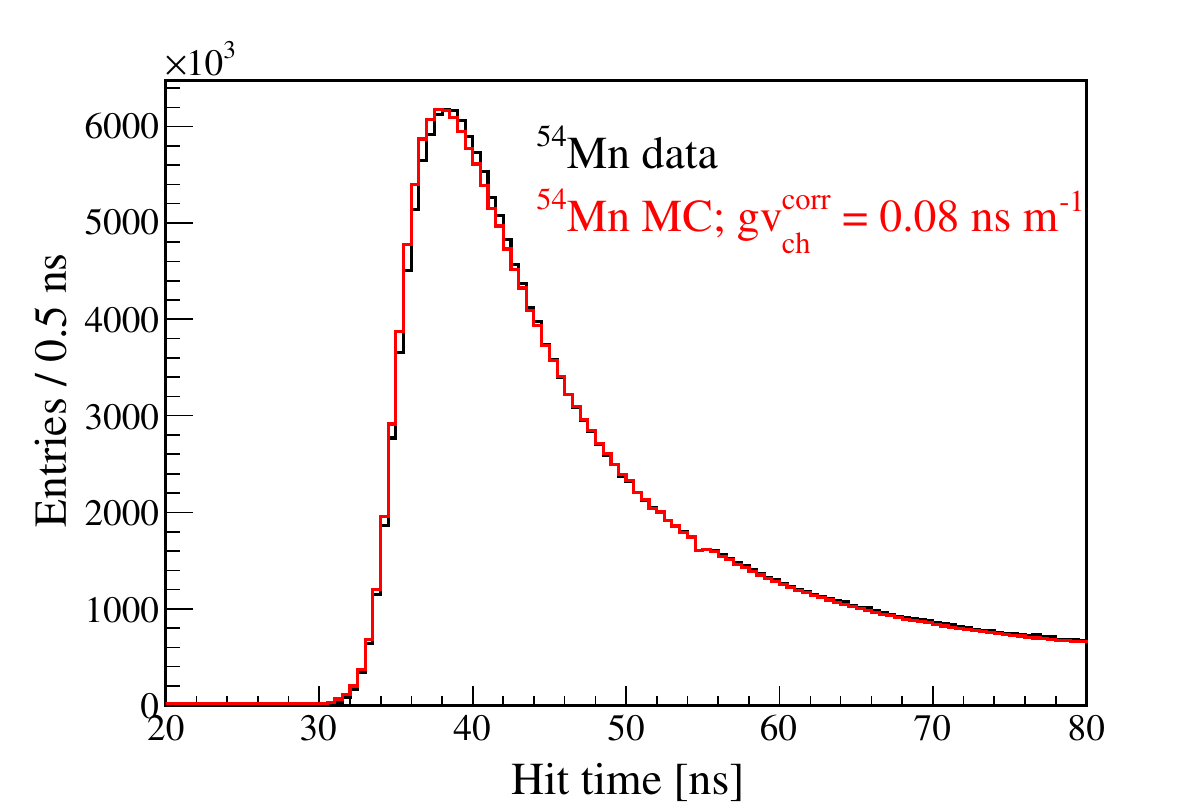}\label{fig:time-gamma}}
    \subfigure[]{\includegraphics[width=0.49\textwidth]{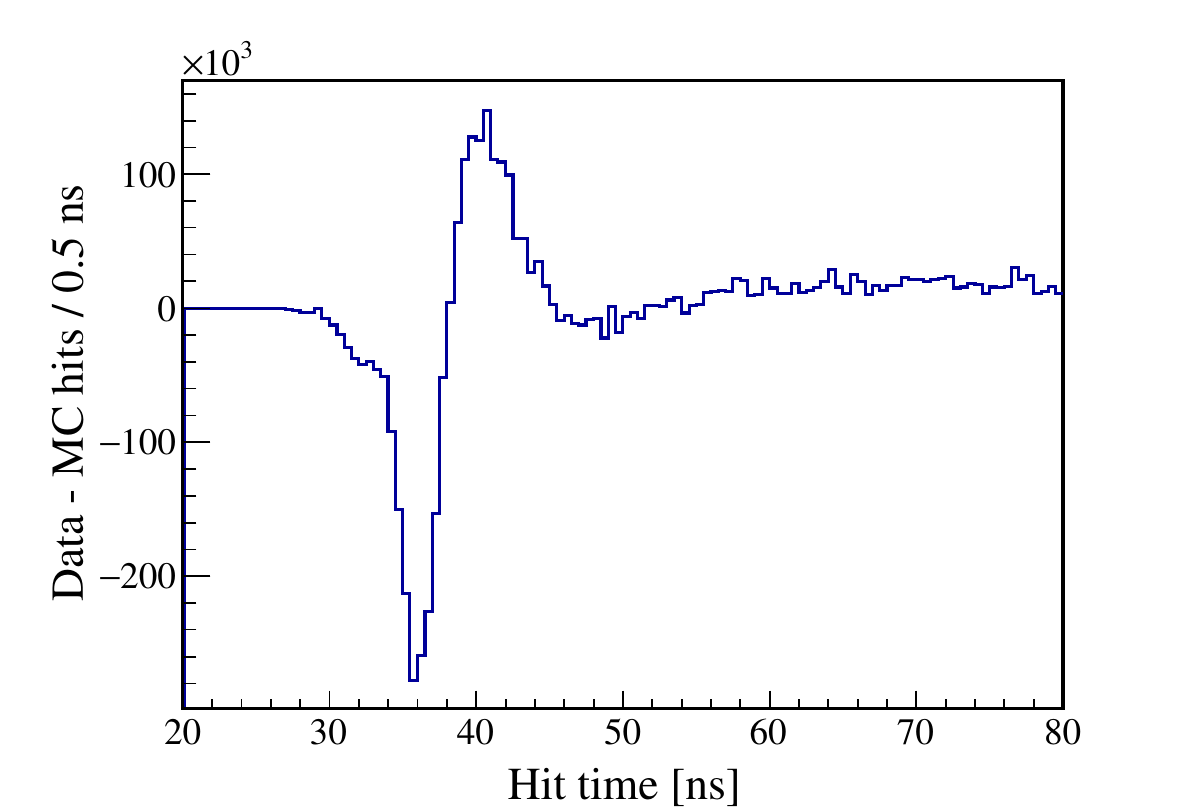}\label{fig:gamma_time_diff}}
    \caption{(a) Time-of-Flight corrected hit times for PMT hits of all events in data (black) and MC (red) of the $^{54}$Mn gamma calibration source. The distributions are normalized to have the same area. (b) The difference between the data and MC time distributions in (a).}
\end{figure*}
Borexino is a high light yield liquid scintillator detector and since systematic differences between data and MC of the sub-dominant Cherenkov photons do not impact the typical spectral analysis, a dedicated Cherenkov calibration has not been necessary. Using the \emph{``N\textsuperscript{th}-hit method''}, the \emph{absolute, average} differences of the PMT hit time distribution between data and MC (such as a constant offset) are effectively resolved. However, the difference between the \emph{relative} time distribution of scintillation and Cherenkov light in data and MC persists. This can affect the amount of Cherenkov photons at early times, and therefore, the relative difference in MC needs to be calibrated.

Since the effective wavelength spectrum of detectable Cherenkov photons in Borexino has never been measured in-situ and the refractive index used in the MC has a finite accuracy, the relative group velocities of scintillation and Cherenkov photons can be different in data and MC. This can essentially increase or decrease the ratio of Cherenkov photons in the first \emph{``N\textsuperscript{th}-hits''} when the Cherenkov light has a significant contribution. Therefore, we rely upon our calibration data performed using $\gamma$ sources placed inside the detector~\cite{BxCalibPaper}, to estimate the systematic group velocity correction $gv_{\mathrm{ch}}^{\mathrm{corr}}$ that needs to be applied in the MC for Cherenkov photons.

The interaction of MeV $\gamma$s in the Borexino LS is dominated by Compton scattering on multiple electrons, before they lose energy and get absorbed by the LS molecules. The electrons in turn excite the LS molecules that emit isotropic scintillation light, as well as an overall relatively small amount of Cherenkov light ($\sim$0.2-0.5\%, depending on energy according to MC). This is depicted in Figure~\ref{fig:cos_delta_sketch}.
As the Compton electrons tend to be scattered more in the forward direction, their initial direction is correlated to the $\gamma$ direction and as such, their Cherenkov photons are also more likely to produce hit patterns correlated to the $\gamma$ direction. Thus using the reconstructed $\gamma$ direction, it is possible to define a correlated angular distribution similar to CID:
\begin{equation}\label{eq:cosdelta_def}
    \cos{\delta}_i = \frac{
    \left(\vec{r}^{\hspace{0.5mm}\mathrm{PMT}}_{i} - \vec{r}_{\mathrm{source}}\right) \cdot
    \left(\vec{r}_{\mathrm{rec}} - \vec{r}_{\mathrm{source}}\right)}
    {|\left(\vec{r}^{\hspace{0.5mm}\mathrm{PMT}}_{i} - \vec{r}_{\mathrm{source}}\right)| |\left(\vec{r}_{\mathrm{rec}} - \vec{r}_{\mathrm{source}}\right)| }.
\end{equation}
Equation~\eqref{eq:cosdelta_def} defines the directional angle $\cos{\delta}$ of the PMT hits used for this calibration, calculated using the position of the PMT that detected the hit $\vec{r}^{\hspace{0.5mm}\mathrm{PMT}}_{i}$ and the reconstructed position of the $\gamma$ event $\vec{r}_{\mathrm{rec}}$. Here $|\left(\vec{r}_{\mathrm{rec}} - \vec{r}_{\mathrm{source}}\right)|$ is the reconstructed gamma direction, based on the reconstructed event position.

The group velocity correction $gv_{\text{ch}}^{\mathrm{corr}}$ is implemented in MC as follows:
\begin{align}\label{eq:gv-corr}
\nonumber
t^{\mathrm{ToF}}_{\mathrm{new}} &=  t^{\mathrm{ToF}}_{\mathrm{old}} - \left(gv_{\text{ch}}^{\mathrm{corr}}\cdot L_{\mathrm{true}}\right)\\
&=  t^{\mathrm{ToF}}_{\mathrm{old}} - \left(\frac{\Delta n_{\mathrm{ch}}}{c}\cdot L_{\mathrm{true}}\right),
\end{align}
where $t_{\mathrm{new}}^{\mathrm{ToF}}$ is the modified hit time of the Cherenkov photons, $t_{\mathrm{old}}^{\mathrm{ToF}}$ is the normal hit time of the Cherenkov photons in MC and $L_{\mathrm{true}}$ is the MC photon track length. The group velocity correction $gv_{\text{ch}}^{\mathrm{corr}}$ has the unit of  ns m$^{-1}$,  and it is an effective parameter that is used to change the relative timing between scintillation and Cherenkov light such that the MC and data have compatible $\cos{\delta}$ distributions. This relation can be further expressed as a function of the effective change in the refractive index at a particular wavelength $\Delta n_{\mathrm{ch}}$. Since the MC is changed at the lowest level where the true origin and track length of each photon are known, the results obtained from the $\gamma$ calibration are applicable also to the MC Cherenkov photons of electrons.

The $gv_{\mathrm{ch}}^{\mathrm{corr}}$ is a sub-nanosecond effect and it depends on the overall timing behavior of the whole detector such as the PMT and electronics timing properties, dark noise, the distribution of active PMTs, and possible other unknown effects. Since the $\gamma$ sources were deployed in mid-2009, this result should be considered reliable only for Phase-I (May 2007\textendash May 2010) of the Borexino experiment. For the later Phase-II (December 2011\textendash May 2016) and Phase-III (July 2016\textendash ongoing), it is not possible to exclude the change or degradation of the overall detector timing properties with a sub-nanosecond precision.

\subsection{Data selection for $\gamma$-sources}

This analysis uses calibration data from $^{54}$Mn and $^{40}$K gamma sources, with $Q$-values of 0.834\,MeV and 1.460\,MeV, respectively~\cite{BxCalibPaper}. The sources were placed at the center of the detector and their position is known with an uncertainty of \SI{1}{cm}, measured by CCD cameras during the calibration campaign \cite{BxCalibPaper}. The chosen energy regions are 290\,$< N_{h} <$\,350 and 480\,$< N_{h} <$\,600 for $^{54}$Mn and $^{40}$K sources, respectively. In addition, a radial cut of $<$0.8\,m with respect to the source position is applied, so as to only select $\gamma$-events arising from the source. 

\subsection{Methods and strategy of analysis}
\begin{figure*}[t!]
    \centering
    \subfigure[]{\includegraphics[width=0.49\textwidth]{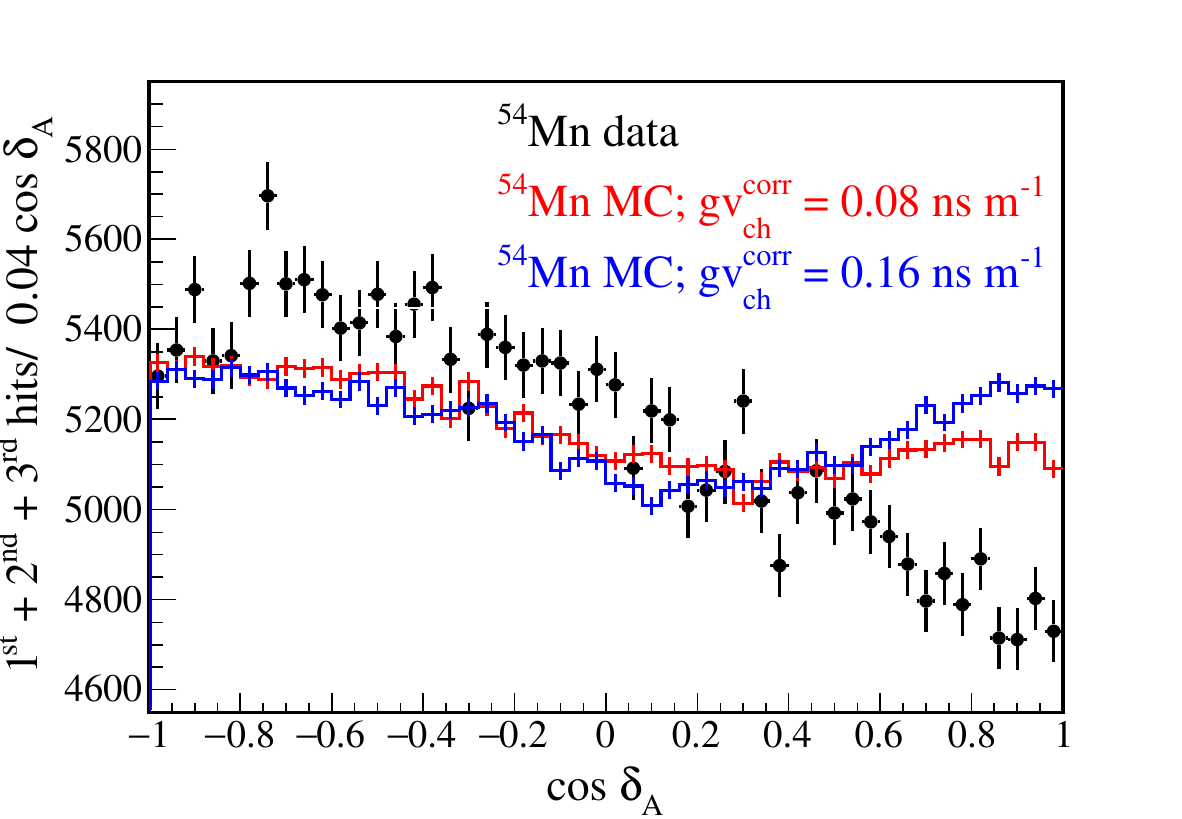}\label{fig:bad_Mn}}
    \subfigure[]{\includegraphics[width=0.49\textwidth]{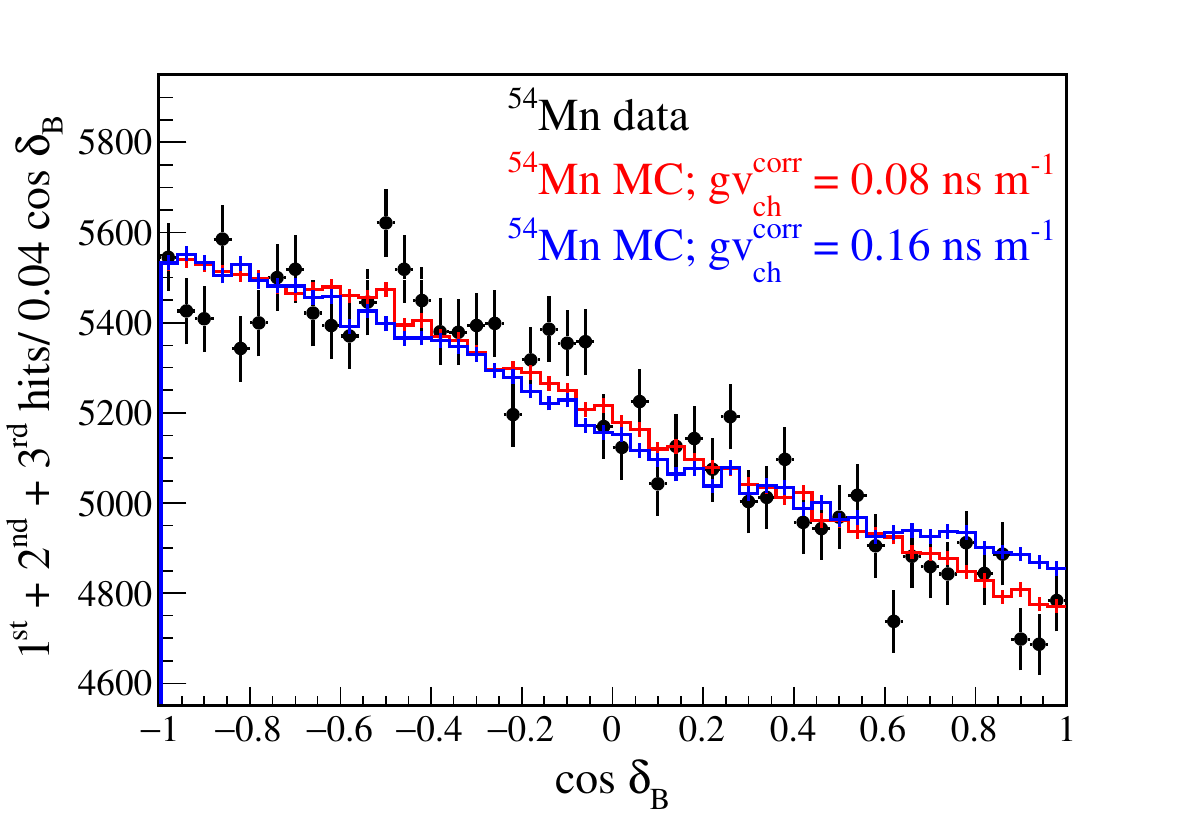}\label{fig:Mn-PDF}}
    \subfigure[]{\includegraphics[width=0.49\textwidth]{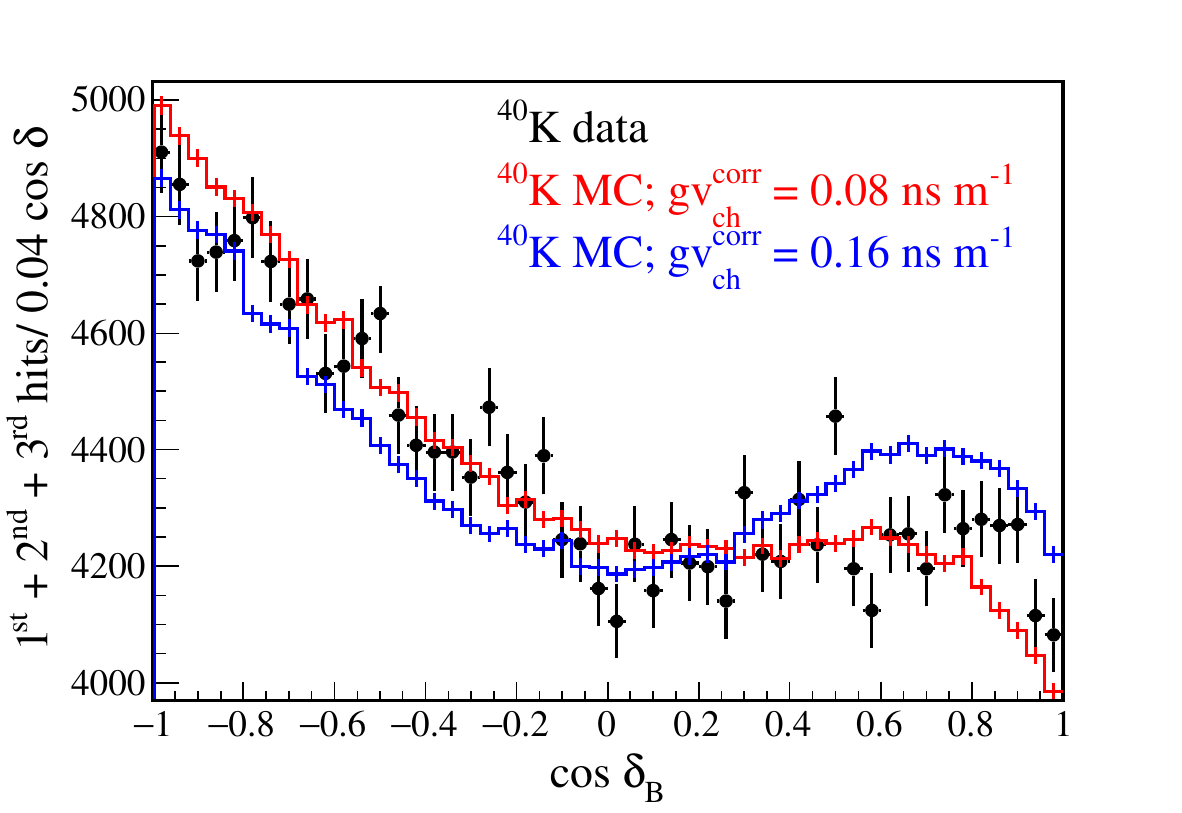}\label{fig:K-PDF}}
    \subfigure[]{\includegraphics[width=0.49\textwidth]{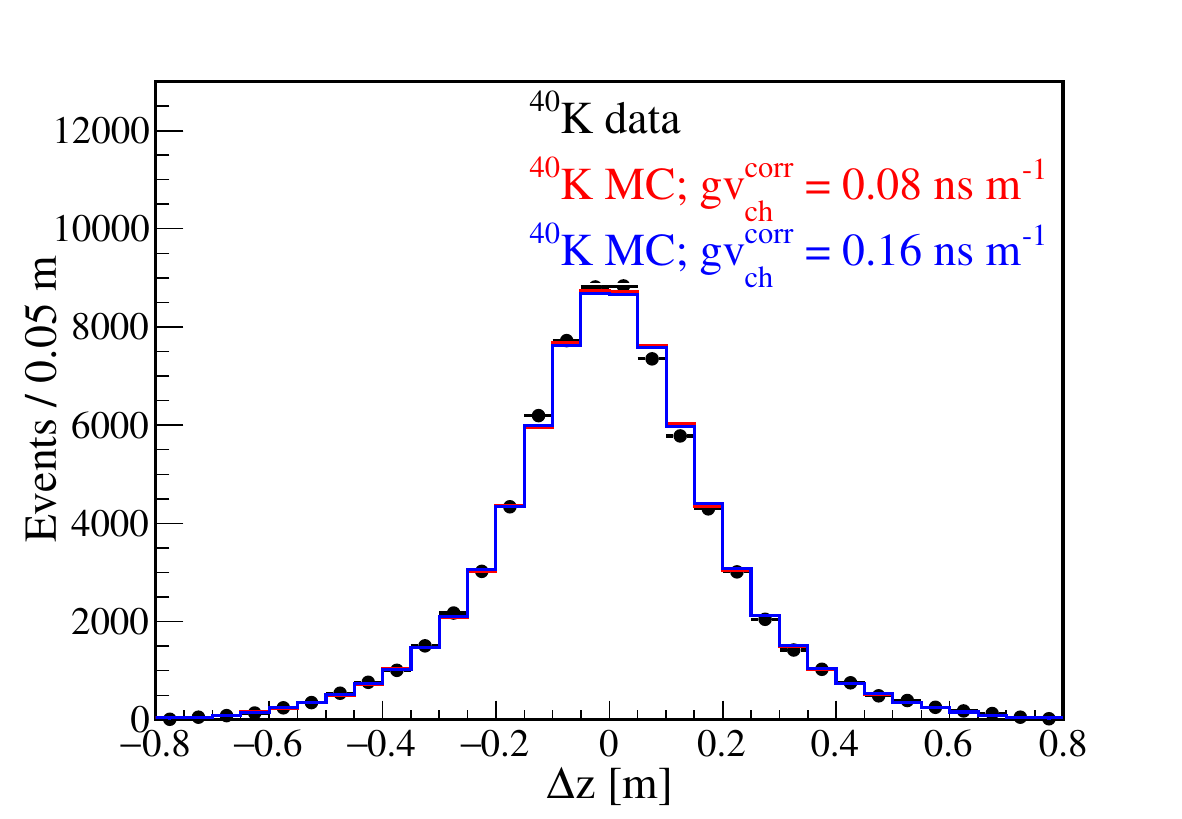}\label{fig:gamma_pos_rec}}
    \caption{Example of the effect of using separate, modified direction reconstruction PDFs, for data and MC $\gamma$ sources for two different values of $gv_{\text{ch}}^{\mathrm{corr}}$: \SI{0.08}{\nano\second\per\meter}(red) , \SI{0.16}{\nano\second\per\meter}(blue).  (a) $\cos{\delta}$ distributions of the $^{54}$Mn source where the direction of data (black) and MC (red, blue) are reconstructed with the same unmodified direction reconstruction PDF. (b) $\cos{\delta}$ distributions of the same $^{54}$Mn source for one scenario where the direction is reconstructed with different, modified PDFs for data and both $gv_{\text{ch}}^{\mathrm{corr}}$ MC simulations. The PDFs are selected to give a good agreement for the $\cos{\delta}$ distributions. (c) $\cos{\delta}$ distributions of the $^{40}$K source with the same combination of direction reconstruction PDFs that are used for $^{54}$Mn in (b). (d) Position reconstruction of data (black) and MC (red, blue) with the modified position reconstruction PDFs that are also used for (b), (c). Both the modification of the PDFs and the Cherenkov group velocity correction $gv_{\text{ch}}^{\mathrm{corr}}$ do not impact the position reconstruction performance, as shown for $^{40}$K.}
    \label{fig:gamma-dir-pdf}
\end{figure*}
The goal of the $\gamma$ Cherenkov calibration is to estimate $gv_{\text{ch}}^{\mathrm{corr}}$ to be used in the solar neutrino analysis in Equation~\eqref{eq:nusol-fit}. This is done by performing a $\chi^{2}$-fit between the $\cos{\delta}$ histograms of data and MC of the $\gamma$ source calibration with $gv_{\text{ch}}^{\mathrm{corr}}$ as a free parameter. The analysis proceeds methodically the same way as the solar neutrino analysis, by sorting the PMT hits of all the selected events in ToF corrected hit time, as described in Section~\ref{sec:analysis}. The cut on the N\textsuperscript{th}-hit has been chosen by studying the $\chi^{2}$ difference between $^{40}$K MC $\cos{\delta}$ distributions with different values of $gv_{\text{ch}}^{\mathrm{corr}}$ (0.08\,ns/m and 0.22\,ns/m). It has been found that the first 3 hits are expected to be the most sensitive to the group velocity correction applied to Cherenkov photons, corresponding to a Cherenkov/scintillation ratio of $\sim$3-12\%, depending on the $gv_{\text{ch}}^{\mathrm{corr}}$.

Ideally, the $\cos{\delta}$ fit can be performed on any of the $\gamma$ sources by leaving the $gv_{\text{ch}}^{\mathrm{corr}}$ as a free parameter. However, unlike the solar analysis, where the direction of the solar neutrinos is well known due to the position of the Sun, here the direction of the $\gamma$ must be reconstructed (see Equation~\ref{eq:cosdelta_def}). This direction reconstruction introduces a systematic difference between data and MC as explained below. Figure~\ref{fig:time-gamma} shows the ToF corrected PMT hit time distributions for $^{54}$Mn data and MC. While they look similar, they are significantly different within their statistics in the first few nanoseconds. The difference between the data and MC time distributions is shown in Figure~\ref{fig:gamma_time_diff}.

The reconstruction of an event position in Borexino is based on the time distribution of the collected photons. The algorithm considers the hit time for each detected PMT hit $t_i$ and the position of the PMT $\vec{r}^{\hspace{0.5mm}\mathrm{PMT}}_{i}$ that detected the hit and subtracts its Time-of-Flight for a possible position $\vec{r}_{\mathrm{rec}}$. Then this photon time distribution is compared with the reference time PDF of the Borexino scintillator. The event position is calculated by maximizing the likelihood $\mathcal{L}(\vec{r}_{\mathrm{rec}},t_0 |\vec{r}^{\hspace{0.5mm}\mathrm{PMT}}_{i})$ that the event occurs at the time $t_0$ in the position $\vec{r}_{\mathrm{rec}}$ given the measured hit space-time pattern $(\vec{r}^{\hspace{0.5mm}\mathrm{PMT}}_{i},t_i)$ \cite{BxCalibPaper}.

This means that the differences of the underlying hit time distributions of data and MC will produce different reconstructed positions for the same true event position. For the difference between the true source position and the reconstructed event position, both data and MC perform on average the same with a resolution of \SI{20}{cm} for the $^{54}$Mn $\gamma$ source. This is different for the direction reconstruction of the 1$\textsuperscript{st}$, 2$\textsuperscript{nd}$, 3$\textsuperscript{rd}$ earliest event hits as they are sensitive to the reconstructed position relative to the PMT position. This effect can be seen in Figure~\ref{fig:bad_Mn}, where the sum of the $\cos{\delta}$ distributions of the 1$\textsuperscript{st}$+ 2$\textsuperscript{nd}$+3$\textsuperscript{rd}$ event hits is shown for data and two different $gv_{\text{ch}}^{\mathrm{corr}}$ MC simulations. It can be seen that there is no $gv_{\text{ch}}^{\mathrm{corr}}$ for which MC agrees with data and thus no $gv_{\text{ch}}^{\mathrm{corr}}$ can be estimated this way. To summarize, the difference in the underlying hit time distributions of data and MC introduces a systematic effect in the direction reconstruction of the $\gamma$ events which must be corrected and its uncertainty must be evaluated. 

The time differences between data and MC can be empirically described as a first Gaussian derivative (Figure~\ref{fig:gamma_time_diff}), and is defined as:
\begin{equation}\label{eq:gamma-gaus}
f(t) = A\cdot(\mu-t)\cdot \exp{\bigg(-\frac{(t-\mu)^2}{\sigma^2}\bigg)},
\end{equation}
where $A$ is the amplitude, $\mu$ is the mean, and $\sigma$ is the standard deviation. This function is then added to the Borexino position reconstruction PDF (which is a function of the hit times) to produce a modified position reconstruction PDF. This is done separately for data and each $gv_{\text{ch}}^{\mathrm{corr}}$ MC. As the direction reconstruction is based entirely on the position reconstruction, from now on these PDFs will only be called direction reconstruction PDFs. The parameters $A_{\mathrm{data}}$, $\mu_{\mathrm{data}}$, $\sigma_{\mathrm{data}}$, $A_{\mathrm{MC}}$, $\mu_{\mathrm{MC}}$, $\sigma_{\mathrm{MC}}$ are selected such that the different modified direction reconstruction PDFs produce $\cos{\delta}$ histograms that are in agreement between $^{54}$Mn data and MC for each $gv_{\text{ch}}^{\mathrm{corr}}$. MC and data $\cos{\delta}$ histograms are considered as agreeable in this analysis, if $\chi^{2}$/ndf$<1.5$, using 50 bins. The same $A_{\mathrm{data}}$, $\mu_{\mathrm{data}}$, $\sigma_{\mathrm{data}}$, $A_{\mathrm{MC}}$, $\mu_{\mathrm{MC}}$, $\sigma_{\mathrm{MC}}$ are then applied to the direction reconstruction PDFs of $^{40}$K (\SI{1.460}{MeV}), which has a higher energy than $^{54}$Mn (\SI{0.834}{MeV}) and thus more Cherenkov photons.

Figure~\ref{fig:gamma-dir-pdf} shows the effect of using one such modified direction reconstruction PDFs on the $\cos{\delta}$ distributions of data and MC. For the $^{54}$Mn source (Figure~\ref{fig:Mn-PDF}), the different $gv_{\text{ch}}^{\mathrm{corr}}$ MC simulations are well in agreement with data and there is no sensitivity left for a fit on $gv_{\text{ch}}^{\mathrm{corr}}$, but the systematic difference (seen in Figure~\ref{fig:bad_Mn}) of the direction reconstruction has been resolved. For the $^{40}$K source (Figure~\ref{fig:K-PDF}), the different $gv_{\text{ch}}^{\mathrm{corr}}$ simulations are now distinguishable, as $^{40}$K has more Cherenkov photons than $^{54}$Mn, and therefore has sensitivity for a $gv_{\text{ch}}^{\mathrm{corr}}$ fit. Figure~\ref{fig:gamma_pos_rec} shows that the group velocity correction $gv_{\text{ch}}^{\mathrm{corr}}$ as well as the slightly modified position reconstruction PDFs do not influence the actual reconstructed position of the event, indicating that this effective correction cannot influence the typical spectral analysis in Borexino. The $gv_{\text{ch}}^{\mathrm{corr}}$ is fitted using a $\chi^{2}$-fit of the $\cos{\delta}$ histograms of $^{40}$K data and MC: 
\begin{equation}\label{eq:gamma-fit}
        \chi^{2}(gv_{\mathrm{ch}}^{\mathrm{corr}}) = \sum \limits_{n=1}^{N}\sum\limits_{i=1}^{I}\frac{\bigg((\cos{\delta})^{D}_{n, i} - (\cos{\delta})^{M}_{n, i}\big(gv_{\mathrm{ch}}^{\mathrm{corr}}\big)\bigg)^{2}}{(\sigma^{D}_{n, i})^{2} + (\sigma^{M}_{n, i})^{2}}.
\end{equation}
The index $n$ runs from 1 to the selected N\textsuperscript{th} hit, where N = 3, as discussed before and the index $i$ runs from 1 to the total number of bins $I$=50 in the range -1\,$< \cos{\delta} <$\,+1. $(\cos{\delta})^{D}_{n, i}$ and $(\cos{\delta})^{M}_{n, i}$ are the $\cos{\delta}$ values for the i\textsuperscript{th} bin of the n\textsuperscript{th} hit of data and MC, respectively, and, $\sigma^{D}_{n,i}$ and $\sigma^{M}_{n, i}$ are their respective statistical errors.

\subsection{Results on the Cherenkov group velocity}

\begin{figure}[t]
    \centering
    \includegraphics[width=0.49\textwidth]{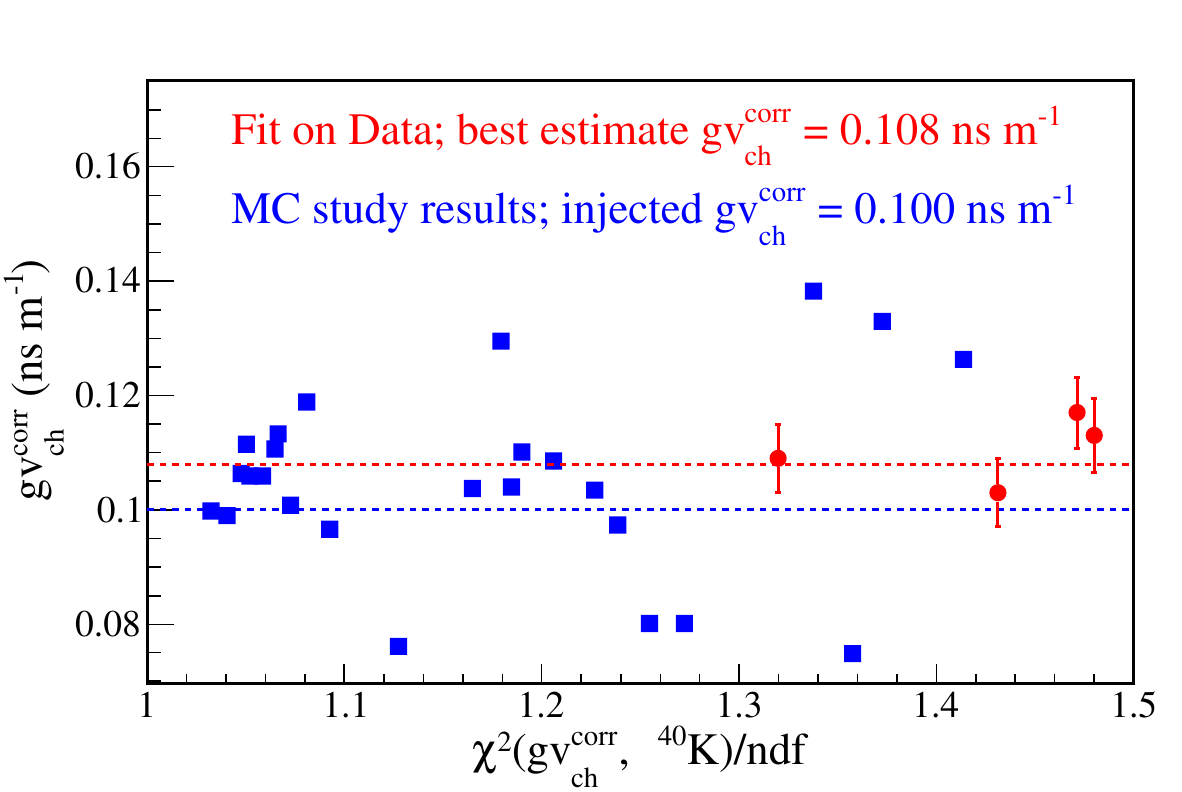}
    \caption{The $gv_{\text{ch}}^{\mathrm{corr}}$ fit results of the gamma Cherenkov calibration as a function of $\chi^2$/ndf of the $^{40}$K $\cos{\delta}$ histograms (Equation~\ref{eq:gamma-fit}). The red data points are the $gv_{\text{ch}}^{\mathrm{corr}}$ estimated from the fit between data and MC, for which the best estimate \SI{0.108}{\ns\per\m} is given by the red dotted line. The blue squares represent the extracted $gv_{\text{ch}}^{\mathrm{corr}}$ from MC studies, falling in the same $\chi^{2}$/ndf space, for an injected $gv_{\text{ch}}^{\mathrm{corr}}$\,=\SI{0.10}{\ns\per\m} represented by the blue dotted line.}
    \label{fig:gamma-res}
\end{figure}

The $gv_{\text{ch}}^{\mathrm{corr}}$ is fitted with the $^{40}$K data and MC using Equation~\eqref{eq:gamma-fit}, after correcting the direction mis-reconstruction using $^{54}$Mn data and MC. The same $^{54}$Mn, $^{40}$K data were analyzed for a number of different sets of direction reconstruction PDFs. Here a \emph{set} describes a number of different direction reconstruction PDFs for data and each $gv_{\text{ch}}^{\mathrm{corr}}$ MC on which an analysis is performed. Each different set of PDFs gives a single fit result $gv_{\text{ch}}^{\mathrm{corr}}$, as well as an overall compatibility of data and MC given by the $\chi^{2}(gv_{\mathrm{ch}}^{\mathrm{corr}})$ in Equation~\eqref{eq:gamma-fit} for $^{54}$Mn and $^{40}$K. The fit results are only considered relevant when the best fit $gv_{\text{ch}}^{\mathrm{corr}}$ fulfills these conditions, for a binning of 50:
\begin{equation} \label{eq:chi2_gamma_condition}
\begin{split}
\chi^{2}(gv_{\mathrm{ch}}^{\mathrm{corr}} ,~ ^{54}\mathrm{Mn})/ndf < 1.5, \\
\chi^{2}(gv_{\mathrm{ch}}^{\mathrm{corr}} ,~ ^{40}\mathrm{K})/ndf < 1.5.
\end{split}
\end{equation}
 The result is shown in Figure~\ref{fig:gamma-res} where the best fit $gv_{\text{ch}}^{\mathrm{corr}}$ is plotted as a function of the $\chi^2$/ndf between the $^{40}$K data and MC $\cos{\delta}$ histograms. The red points are the fit results on data for different sets of direction reconstruction PDFs, with the red dotted line showing the best estimate at \SI{0.108}{\ns\per\m}. The blue squares are the fit results of a MC study with an injected value of \SI{0.10}{\ns\per\m} shown with the dotted blue line.
 
 Different direction reconstruction PDFs used can result in different fitted $gv_{\text{ch}}^{\mathrm{corr}}$ values and this systematic uncertainty has been evaluated in a MC study. The MC study has been performed in the same way as the data analysis for many different sets of direction reconstruction PDFs to estimate the possible offset between the injected and extracted values. It can be seen in Figure~\ref{fig:gamma-res} that for an injected value $gv_{\mathrm{ch}}^{\mathrm{corr}}=\SI{0.10}{\ns\per\m}$, different values of $gv_{\text{ch}}^{\mathrm{corr}}$ are extracted. Similar to the data analysis, the $gv_{\mathrm{ch}}^{\mathrm{corr}}$ found by the fit are only considered relevant in the $\chi^{2}$ space defined in Equation~\ref{eq:chi2_gamma_condition}. The systematic uncertainty introduced by the difference in the direction reconstruction of data and MC can then be estimated from the MC study as the largest offset relative to the injected $gv_{\text{ch}}^{\mathrm{corr}}$ value, which is $\Delta gv_{\text{ch}}^{\mathrm{corr}}=\SI{0.039}{\ns\per\m}$. This MC study has been performed also for an injected $gv_{\mathrm{ch}}^{\mathrm{corr}}$ value of \SI{0.16}{\ns\per\m}, which gives consistent results.
 
 Further systematic studies have been performed, but the uncertainty due to the direction reconstruction is dominant. Due to the possible presence of events other than $\gamma$-events from the source, the systematic influence of the $N_{h}$ (energy) cut is estimated to be \SI{0.004}{\ns\per\m}, after performing the analysis with different $N_{h}$ cuts. The exact choice of the N\textsuperscript{th} hit considered in the analysis has a systematic uncertainty of \SI{0.006}{\ns\per\m}. The statistical uncertainty measured for the data fit is also in agreement with the expected uncertainty from the toy MC study. Taking the fit value with the best $\chi^{2}(gv_{\mathrm{ch}}^{\mathrm{corr}}, ^{40}K)$ as the result of the Cherenkov group velocity correction gives:
\begin{equation}
 gv_{\mathrm{ch}}^{\mathrm{corr}} = \SI[parse-numbers=false]{0.108\pm0.006(stat).\pm0.039(syst.)}{\ns\per\meter}.\nonumber
\end{equation}
This group velocity correction is an \emph{effective} correction, as it changes only the timing of Cherenkov photons relative to that of scintillation, in such a way that there is an agreement between the directional data and MC $\cos{\delta}$ distributions of the $^{54}$Mn and $^{40}$K gamma sources. This can be further expressed as a change in the refractive index, according to Equation~\eqref{eq:gv-corr}, $\Delta n_{\mathrm{ch}} = 0.032 \pm 0.012$. This is only a 2\% correction, considering the refractive index of $\approx$1.55 @ 400\,nm~\cite{Borex-mc}.

It can be concluded that the use of $\gamma$ sources for Cherenkov calibration is not optimal since it gives a relative systematic uncertainty of 36\% on the group velocity correction. However, it is still possible to measure the solar neutrino signal (Section~\ref{sec:results}), even with this relatively large uncertainty on $gv_{\mathrm{ch}}^{\mathrm{corr}}$.

\section{Systematic uncertainties}
\label{sec:sys-others}

The major contributions to the uncertainty on the final directionality measurement arises from the bias of the position reconstruction of electrons ($\Delta_{\mathrm{dir}}$, Section~\ref{sec:analysis}), and the group velocity estimation ($gv_{\mathrm{ch}}^{\mathrm{corr}}$, Section~\ref{sec:gamma_calib}), which are included in the final fit as a free nuisance parameter and as a nuisance parameter with a Gaussian pull term, respectively. The other, smaller systematic uncertainties can be divided into two categories: (1) The systematic uncertainties on the directionality measurement ($N_{\mathrm{solar}-\nu}$). (2) The systematic uncertainties arising from the conversion of the total number of solar neutrinos ($N_{\mathrm{solar}-\nu}$) into the $^{7}$Be interaction rate ($R(^{7}$Be)) in the detector. The former category includes: the chosen cut on the N\textsuperscript{th}-hit, the method of selection of PMTs used in the analysis, and the choice of histogram binning. Moreover, there are many reasons why the background does not have a perfectly flat $\cos{\alpha}$ distribution. These effects were studied individually and it has been concluded that they do not contribute to the systematic uncertainties. The latter category of uncertainties on $R(^{7}$Be) additionally includes: the uncertainty on the exposure, the uncertainty on the efficiency of the MLP variable used for $\alpha/\beta$ discrimination, and the uncertainty due to the theoretical predictions of CNO and \emph{pep} neutrino rates which are used for the conversion. The uncertainties on the energy and trigger efficiencies used are negligible. All the systematic effects are summarized in Table~\ref{tab:summary_sys}.

\paragraph{Choice of N\textsuperscript{th}
Hit}
The analysis is performed on the first two hits of the selected events as described in Section~\ref{sec:analysis}. The selection of this N\textsuperscript{th}-hit bears a systematic uncertainty and has been estimated to be 4.8\% after performing the fit on different N\textsuperscript{th}-Hit cuts (2,3,4).
\begin{table}[t!]
	\centering
	\caption{\label{tab:summary_sys} Summary of the different sources of systematic uncertainty for the number of solar neutrinos $N_{\mathrm{solar}-\nu}$ and the $^{7}$Be interaction rate  $R(^{7}$Be) in the ROI. Different contributions are summed up as uncorrelated.} \vskip 2pt
	\begin{tabular*}{\columnwidth}{l @{\hskip 60pt} c}
		\hline \hline
		Source & Uncertainty [\%] \TstrutLarge \BstrutLarge \\
		\hline
		Choice of N\textsuperscript{th} Hit & 4.8 \Bstrut \\
		Selection of PMTs & 5.9 \Bstrut \\
		Choice of histogram binning & 4.2 \Bstrut\\
		\hline
		Total for $N_{\mathrm{solar}-\nu}$ & 8.7 \TstrutLarge\BstrutLarge \\
		\hline
		Exposure & 4.6 \Bstrut \\
		MLP variable & 1.0 \Bstrut \\
		CNO and \emph{pep} rates & $^{+2.3}_{-1.2}$ \Bstrut \\
		\hline
		Total for $R(^{7}$Be) &  $^{+10.1}_{-10.0}$ \TstrutLarge\BstrutLarge \\
		\hline
		\hline
	\end{tabular*}
\end{table}
\paragraph{Choice of histogram binning}
As the analysis is performed on a binned data sample, the choice of the bin-width has been studied using a MC study. When the number of bins becomes too low, the expected uncertainty increases as  usable information is smeared out. For a fixed injected $N_{\mathrm{solar}-\nu}$, the standard deviation of the distribution of the extracted $N_{\mathrm{solar}-\nu}$ showed a steady decrease until 20 bins, followed by a stable precision between 20 and 80 bins. The CID analysis has been then performed with 30, 40, 60, 120 bins, resulting in a systematic uncertainty of 4.2\%.

\paragraph{Selection of PMTs}
As mentioned in Section~\ref{sec:data-sel}, the distribution of PMTs influence the $\cos{\alpha}$ distribution. Therefore, it has to be guaranteed that only PMTs with the same qualitative first hit time behavior are used in MC and data. So, we apply a selection based on the number of hits contributed by each PMT, such that they are statistical compatible between data and MC. These systematic differences are only present in the first few hits of the event, and are usually not of concern when all the the hits of the event are used for the other analyses in Borexino. The method of selection gives a systematic uncertainty of 5.9\%.

\paragraph{Effects on the background distribution}
The $\cos{\alpha}$ distribution of the background is not entirely flat, although it has no correlation to the Sun's position. This effect is seen in the distribution of the $^{210}$Po $\alpha$ background in data (Figure~\ref{fig:cos_alpha_data}) and the $^{210}$Bi, $\beta$ MC (Figure~\ref{fig:cos_alpha_mc}). It is due to multiple effects: (1) non-uniform background distribution, (2) non-uniform distribution of PMTs, (3) number of live PMTs, (4) non-isotropic distribution of the Sun's position with respect to Borexino co-ordinates, and (5) an asymmetrical fiducial volume cut. These effects were studied using a toy MC framework by first assuming ideal detector conditions, and then applying one effect at a time. All these effects (except for the non-uniform background distribution) are perfectly reproduced by the full MC used in the final fit and they are consistent with the estimation from the toy MC study. Therefore, they do not contribute to the systematic uncertainty.
The effect of the asymmetrical fiducial volume cut, though reproduced by the MC, has been removed by employing a spherical fiducial volume as previously explained in Section~\ref{sec:data-sel}.
The maximum effect due to the possible non-uniformity of background events, which is not reproduced by the MC, has been observed at the level of 0.1\% for Phase-I, thus negligible.

\paragraph{Exposure}
The uncertainty on the exposure is dominated by the precision of the position reconstruction, based on which we select events inside our fiducial volume described in Section~\ref{sec:data-sel}. This effect has been studied thoroughly using various sources during the calibration campaign~\cite{BxCalibPaper}. The maximum uncertainty on the position of the fiducial volume has been found to be 5\,cm~\cite{Geo-longpaper}. Considering a nominal spherical fiducial volume of 3.3\,m, this results in an uncertainty of 4.6\% on the exposure, and therefore also on $R(^{7}$Be).

\paragraph{MLP variable}
The $\alpha/\beta$ discrimination variable we employ has an average efficiency of (99.5$\pm$1.0)\%, where the uncertainty is conservative and takes into account the change with respect to time and also the method used for the efficiency estimation. Therefore, we consider this as a systematic uncertainty on $R(^{7}$Be).

\paragraph{CNO and pep rates} 
The conversion of $N_{\mathrm{solar}-\nu}$ to $R(^{7}$Be) requires fixing the CNO and \emph{pep} solar neutrino contributions in the ROI. The interaction rates differ for high metallicity (HZ) and low metallicity (LZ) predictions of the Standard Solar Model (SSM)~\cite{Vinyoles:2016djt, CNOsens}. In this analysis, we assume HZ predictions and take the difference between HZ and LZ predictions as a systematic uncertainty (+2.0\%). In addition, there are also uncertainties arising from the theoretical predictions of the CNO and \emph{pep} rates which results in a 1.2\% relative uncertainty on the number of $^{7}$Be neutrinos. Therefore, summing these individual uncertainties in quadrature, the total uncertainty on $R(^{7}$Be) due to the assumptions of the CNO and \emph{pep} rates is $^{+2.3}_{-1.2}$\%. The systematic uncertainty arising from the different $\cos{\alpha}$ shapes of the CNO and \emph{pep} solar neutrinos in the ROI, according to Equation~\eqref{eq:scatter_angle}, has been studied using Borexino's MC simulation and has been found to be negligible. This means that for the Borexino detector the CID method alone is not able to differentiate between $^{7}$Be, CNO and \emph{pep} neutrinos in the ROI. Only the sum of $^{7}$Be + CNO + \emph{pep} neutrinos can be inferred from the measured $\cos{\alpha}$ distribution.
\\

The total systematic uncertainty on the directional measurement of solar neutrinos ($N_{\mathrm{solar}-\nu}$) and the $^{7}$Be interaction rate are 8.7\%, and $^{+10.1}_{-10.0}$\%, respectively, where the individual uncertainties are summed in quadrature as uncorrelated.

\section{Results}
\label{sec:results}

\begin{figure*}[t!]
    \centering
    \subfigure[]{\includegraphics[width=0.49\textwidth]{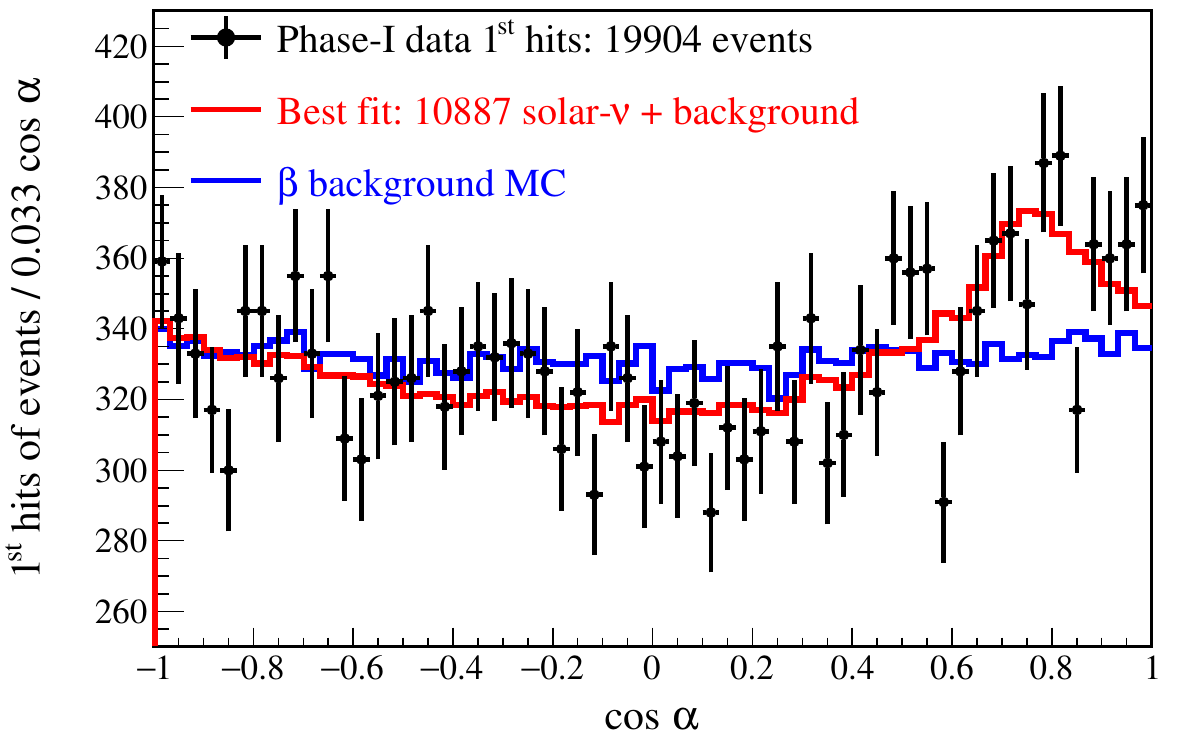}    \label{fig:bestfit_dataMC_1}}
    \subfigure[]{\includegraphics[width=0.49\textwidth]{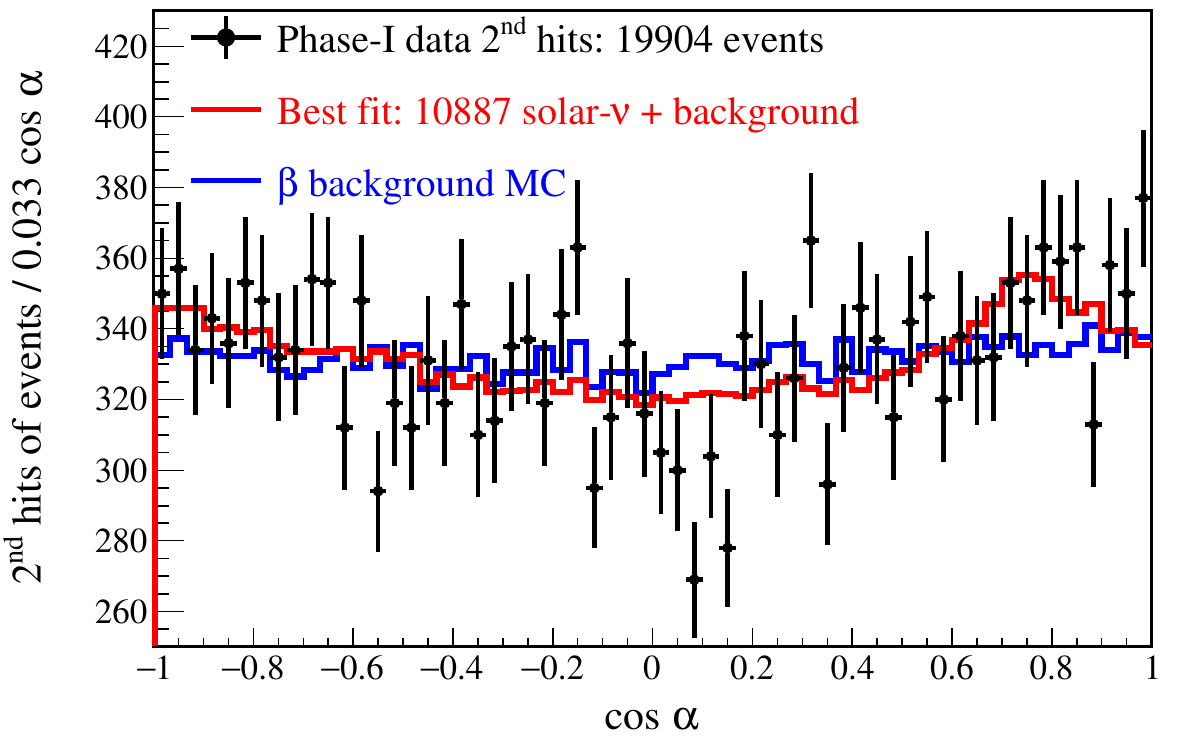}    \label{fig:bestfit_dataMC_2}}
    \subfigure[]{\includegraphics[width=0.49\textwidth]{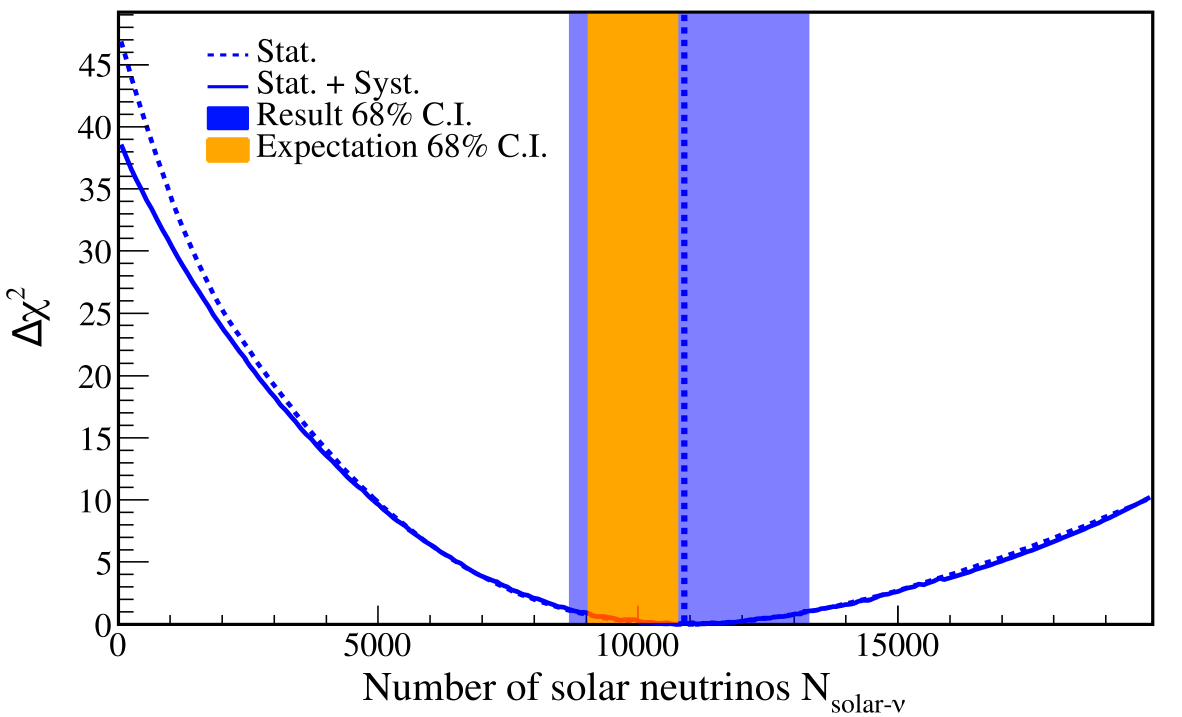}\label{fig:chi2_sigtotot}}
    \caption{The $\cos{\alpha}$ distributions of the first (a) and second (b) hits of all the selected events (black points) compared with the best fit curve (red) for the resulting
    number of solar neutrinos $N_{\mathrm{solar}-\nu}$ plus background, as reported in Section~\ref{sec:results}. All histograms are normalized to the data statistics. It can be seen that the data points cannot be explained by the background-only hypothesis (blue). (c) $\Delta\chi^{2}$ profiles of the first and second hits from the fit as a function of $N_{\mathrm{solar}-\nu}$ with (blue solid curve) and without (blue dotted curve) the systematic uncertainty. The no-neutrino signal hypothesis (pure background, $N_{\mathrm{solar}-\nu}$=0) can be rejected with $\Delta\chi^{2}>25$,  $>5\sigma$.  The 68\%\,CI from the $\Delta\chi^{2}$ profile gives $N_{\mathrm{solar}-\nu}$ = 10887$^{+2386}_{-2103} (\mathrm{stat.})\pm 947 (\mathrm{syst.})$, with a $\chi^{2}/ndf$ = 124.6/117. This is represented by the blue shaded band and the best fit value is shown as a vertical blue dotted line. The 68\%\,CI of the the solar neutrino signal expected based on the Standard Solar Model (SSM) predictions~\cite{CNOsens} is shown as an orange band.}
    \label{fig:results}
\end{figure*}
This section describes the results of the \emph{Correlated and Integrated Directionality} (CID) analysis. In Phase-I of the Borexino experiment, which ran from May 16\textsuperscript{th}, 2007, to May 8\textsuperscript{th}, 2010 corresponding to \SI{740.7}{d} of data acquisition, 19904 events passed the data selection cuts described in Section~\ref{sec:data-sel}. The shapes of the $\cos{\alpha}$ distributions of the selected data and MC events have been discussed in detail in Section~\ref{sec:analysis}. This data is used to give a measurement on the number of solar neutrinos using the CID method, since we are able to calibrate the effective group velocity correction for only Phase-I. The remaining Phase-II and Phase-III data corresponding to livetimes of 1291.5\,d and 1072\,d, respectively, are used to give the exclusion of the no-neutrino hypothesis using CID.

In Phase-I, the $\chi^{2}$-fit described in Section~\ref{sec:analysis} is performed on the first and second PMT hits of all the selected events to obtain the number of solar neutrinos $N_{\mathrm{solar}-\nu}$, which consists of $^{7}$Be+\emph{pep}+CNO neutrino events as described in Section~\ref{sec:data-sel}. Figure~\ref{fig:results} shows the results of the CID analysis.
The best $N_{\mathrm{solar}-\nu}$\,+\,$\beta$-background value from the fit (red) shows a $\cos{\alpha}$ distribution that is well in agreement with the first and second hits of the selected data events (black), while the pure background curve (blue) is incompatible with data.
Figure~\ref{fig:chi2_sigtotot} shows the $\Delta\chi^2$ profile (dotted blue curve) as a function of $N_{\mathrm{solar}-\nu}$ for the first two hits of each event. The $\chi^{2}$ profile has been further smeared with the systematic uncertainty of 8.7\% described in Section~\ref{sec:sys-others}. The smeared $\Delta\chi^2$ profile is shown as a solid blue curve in Figure~\ref{fig:chi2_sigtotot}. For a $\cos{\alpha}$ histogram with 60 bins, the combined $\chi^2/ndf$ of the first and second hits is $\chi^2/ndf = 124.6/117$, ($p\text{-value}=0.30$). The result of the CID analysis in Borexino is then the measurement of the number of solar neutrinos $N_{\mathrm{solar}-\nu}$ present in the ROI:
 \begin{equation}
      N_{\mathrm{solar}-\nu} = 10887^{+2386}_{-2103} (\mathrm{stat.})\pm 947 (\mathrm{syst.}). \nonumber
\end{equation}
The 68\% CI of the result corresponding to $\Delta\chi^2=1$ is shown as a blue band in Figure~\ref{fig:chi2_sigtotot}. The expected total number of solar neutrinos in the ROI can be calculated using the SSM predictions of $^{7}$Be, CNO, and \emph{pep} solar neutrinos (Table I in ~\cite{CNOsens}). In this calculation, the high metallicity (HZ) prediction has been considered for the central value, and the difference between the high metallicity (HZ) and low metallicity (LZ) predictions is considered as a systematic uncertainty. The expectation on $N_{\mathrm{solar}-\nu}$ is then 10187$_{-1127}^{+541}$ events. The 68\% CI of the expectation is shown as an orange band in Figure~\ref{fig:chi2_sigtotot}. 

The $\cos{\alpha}$ distribution of the background is not influenced by the nuisance parameters $gv_{\text{ch}}^{\mathrm{corr}}$ and $\Delta r_{\mathrm{dir}}$, since the Cherenkov light of the background events is not correlated to the position of the Sun. This is the reason for the asymmetry of the $\Delta\chi^2$ profile in Figure~\ref{fig:chi2_sigtotot}, where the profile becomes steeper for a larger background contribution. The pure background hypothesis can be excluded with $\Delta\chi^2$\,$>$\,25, which corresponds to a $>$5$\sigma$ detection of sub-MeV solar neutrinos using their directional Cherenkov light with the CID method.

The statistical uncertainty (+2386, -2103) on the final measurement is the combination of the actual statistics of data and the effect of the nuisance parameters in the fit. The expected statistical uncertainty for fixed nuisance parameters, is found to be $\pm$1523 events, using MC studies. The contribution of uncertainty from the lack of a dedicated $e^{-}$ Cherenkov calibration can then be estimated as (+1837, -1450) events.

The measured $N_{\mathrm{solar}-\nu}$ from Phase-I can be further converted into the $^{7}$Be interaction rate $R(^{7}\mathrm{Be})$
in Borexino, after fixing the contributions of the CNO and \emph{pep} solar neutrinos from the SSM predictions~\cite{CNOsens}, and using the detection efficiency of the three contributions in the selected ROI. Additional systematic uncertainties due the detection efficiency, exposure, and the difference between the HZ and LZ predictions of CNO and \emph{pep} neutrinos described in Section~\ref{sec:sys-others} are also considered. This results in a $^{7}$Be interaction rate of:
\begin{equation}
    R(^{7}\mathrm{Be}) =51.6^{+13.9}_{-12.5}\, (\mathrm{stat.+syst.})\,\mathrm{cpd/100t }. \nonumber
\end{equation}
\begin{figure}[t!]
    \centering
    \includegraphics[width=0.49\textwidth]{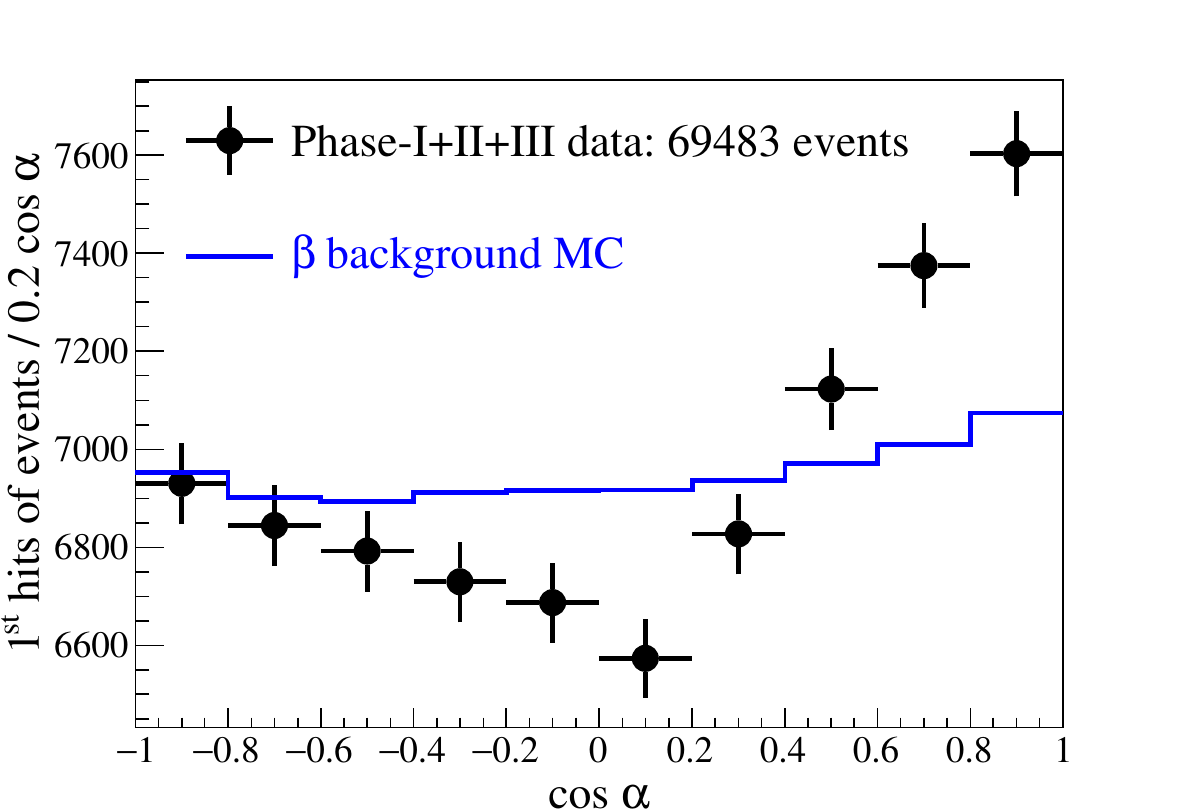}
    \caption{The $\cos{\alpha}$ distribution of Phase I + Phase II + Phase III data events, compared with the $\beta$ background MC PDF normalized to the statistics of data. It can be seen that the data cannot be explained with a background-only hypothesis.}
    \label{fig:phase_1_2_3_2st_data_back}
\end{figure}
In addition to the directional measurement performed with Phase-I, we have tested the background-only hypothesis on the summed data of all the three Borexino Phases. It can be seen from Figure~\ref{fig:phase_1_2_3_2st_data_back} that the first hits of the data cannot be explained with a pure background MC PDF. As Cherenkov light from the background is uncorrelated to the solar neutrino direction, the comparison of data with pure background does not depend on a Cherenkov calibration. Consequently, the detection of a neutrino signal using the Cherenkov light of their recoil electrons is possible for all three Borexino Phases. The $\chi^{2}$ between the data and the MC $\beta$ background PDF gives a $p\text{-value}<3\times10^{-7}$ ($\chi^{2}$/ndf = 95.5/9 for 10 Bins). This result shows that the background-only hypothesis is incompatible with our data with a significance $>$5$\sigma$, which corresponds to a detection of solar neutrinos only using their directionality, without any Cherenkov calibration. 

The directional measurement of solar neutrinos is possible only for Phase-I data, since the gamma Cherenkov calibration is performed using the data from this period. A dependency of the calibration results on the overall detector time response cannot be excluded for later Phases. While the later Phases still show a statistically significant amount of directional information, these data cannot be easily used for a measurement of solar neutrinos without a further Cherenkov calibration.

\section{Conclusions}
\label{sec:conclusion}

In this work, Borexino has provided the first directionality measurement of sub-MeV solar neutrinos in a liquid scintillator detector, through the so-called \emph{Correlated and Integrated Directionality} (CID) method. The CID method produces an angular distribution by correlating the direction of the first few PMT hits of each event to the known solar direction and then integrating these angles over all the selected events. The number of solar neutrinos is then statistically inferred from the contribution of Cherenkov photons correlated to the position of the Sun.

For Phase-I of the Borexino experiment, where we are able to calibrate the effective group velocity correction for Cherenkov photons, the no-neutrino hypothesis has been excluded with a significance greater than 5$\sigma$, purely based on their direction. The number of $^{7}$Be+\emph{pep}+CNO solar neutrino events measured with CID is \emph{N$_{\mathrm{CID}}$} = $10887_{-2103}^{+2386} (\mathrm{stat.}) \pm 947 (\mathrm{syst.})$, considering a total of 19904 events in the ROI. The ROI for this analysis has been selected according to the $N_{h}^{\mathrm{geo}}$ energy spectrum of scintillation light to maximize the expected number of neutrinos over the square-root of background and corresponds to the $^{7}$Be edge. The expected number of solar neutrinos in the ROI is \emph{N$_{\mathrm{SSM}}$} = 10187$_{-1127}^{+541}$ according to the SSM~\cite{CNOsens}, where the uncertainty includes the difference between LZ and HZ models. The CID measurement is well in agreement with the SSM. 

The $^{7}$Be interaction rate in Borexino R($^{7}$Be)$_{\mathrm{CID}}$ has been extracted from the CID measurement R($^{7}\mathrm{Be})_{\mathrm{CID}}$=51.6$^{+13.9}_{-12.5}$\,cpd/\SI{100}{t}, after fixing the \emph{pep} and CNO neutrino rates to their SSM predictions~\cite{Vinyoles:2016djt}, using. This $^{7}$Be rate is also well in agreement with the results of the Phase-I spectroscopy R($^{7}$Be)=47.87$\pm$2.28~cpd/\SI{100}{t}\footnote{This corresponds to the measurement of the 0.862\,MeV mono-energetic line given in~\cite{phase1-nusol}, after also summing the contribution from the 0.384\,MeV line.} and the SSM predictions~\cite{CNOsens}.

In addition, we have also used the combined data from all the three Phases of Borexino to test the background-only hypothesis.  It has been shown that the background-only hypothesis is incompatible with our data with a significance $>$5$\sigma$, confirming the presence of directionality of sub-MeV solar neutrinos in a liquid scintillator detector, even without any calibration of Cherenkov photons. 

The successful measurement of solar neutrinos in a high light yield liquid scintillator detector using only the fit of the directional distribution provided by Cherenkov light and no fit of the energy spectrum is an important proof of principle for the CID method presented here. Thus, this method can be developed further for a joint analysis with a typical spectral fit.

Future solar neutrino LS experiments can readily benefit from the CID method even without specialized hardware or LS mixtures for the separation of Cherenkov and scintillation light. Thus it is highly recommended to perform a dedicated $e^{-}$ Cherenkov calibration for this, even if an event-by-event direction reconstruction based on Cherenkov light is expected to be not possible.

\section*{Acknowledgements}
We acknowledge the generous hospitality and support of the Laboratori Nazionali del Gran Sasso (Italy). The Borexino program is made possible by funding from Istituto Nazionale di Fisica Nucleare (INFN) (Italy), National Science Foundation (NSF) (USA), Deutsche Forschungsgemeinschaft (DFG), Cluster of Excellence PRISMA+ (Project ID 39083149), and recruitment initiative of Helmholtz-Gemeinschaft (HGF) (Germany), Russian Foundation for Basic Research (RFBR) (Grants No. 19-02-00097A) and Russian Science Foundation (RSF) (Grant No. 21-12-00063) (Russia), and Narodowe Centrum Nauki (NCN) (Grant No. UMO 2017/26/M/ST2/00915) (Poland). We gratefully acknowledge the computing services of Bologna INFN-CNAF data centre and U-Lite Computing Center and Network Service at LNGS (Italy).

%

\end{document}